\documentclass[a4paper,12pt]{article}

\usepackage{authblk}
\usepackage{amsmath,amssymb,amsfonts,amsthm,mathrsfs}
\usepackage{hyperref}
\usepackage{comment}
\usepackage{color}

\def\be{\begin{equation}}
\def\ee{\end{equation}}
\def\ba{\begin{eqnarray}}
\def\ea{\end{eqnarray}}
\def\bi{\begin{itemize}}
\def\ei{\end{itemize}}
\def\nn{\nonumber}

\def\bra{\langle}
\def\ket{\rangle}
\def\L{\mathcal{L}}
\def\out{{\rm out}}
\def\in{{\rm in}}
\def\O{\Omega}
\def\xh{\hat{x}}

\def\e{\varepsilon}
\def\qh{\hat{q}}

\def\wb{\bar{w}}
\def\zb{\bar{z}}
\def\w{\omega}

\def\scri{\mathcal{I}}

\def\I{\mathcal{I}}

\def\phitwo{\overset{(-2)}{\varphi}}

\def\Q{\mathcal{Q}}
\def\Qs{\Q^{\rm soft}}
\def\Qh{\Q^{\rm hard}}
\def\rhos{\rho^{\rm soft}}
\def\rhoh{\rho^{\rm hard}}

\def\sint{\textstyle{\int}}
\def\Y{\mathbb{Y}}
\def\Stwo{S^{(2)}}
\def\phid{\dot{\phi}}
\def\tf{\rm TF}
\def\S{\mathcal{S}}
\def\rhot{\rho_{T}}

\def\T{\mathcal{T}}
\def\Tthree{\overset{-3 }{\T}}
\def\Ttwo{\overset{-2 }{\T}}
\def\phitwo{\overset{(-2)}{\varphi}}

\def\hb{\bar{h}}
\def\rhorlr{\, {\overset{\; 1 \ln }{\rho}}}
\def\rhor{\overset{\; 1 }{\rho}}
\def\rholr{\, \overset{\; \ln }{\rho}}
\def\rhoo{\overset{ 0}{\rho}}
\def\xitwor{\overset{(2)}{\xi^r}}
\def\xioner{\overset{(1)}{\xi^r}}
\def\xioneu{\overset{(1)}{\xi^u}}
\def\xioneA{\overset{(1)}{\xi^A}}
\def\xioneB{\overset{(1)}{\xi^B}}
\def\xiou{\overset{(0)}{\xi^u}}
\def\xior{\overset{(0)}{\xi^r}}
\def\xitwou{\overset{(2)}{\xi^u}}
\def\xioA{\overset{(0)}{\xi^A}}
\def\xioB{\overset{(0)}{\xi^B}}
\def\xitwot{\overset{(2)}{\xi^t}}
\def\xionet{\overset{(1)}{\xi^t}}
\def\xiot{\overset{(0)}{\xi^t}}

\def\C{\mathscr{C}}

\def\C{\mathscr{C}}

\def\E{\mathcal{E}}

\def\bo{\mathring{\beta}}
\def\B{\mathcal{B}}

\def\grav{\text{grav}}
\def\matt{\text{matt}}
\def\lin{\text{lin}}
\def\soft{\text{soft}}
\def\hard{\text{hard}}

\def\stf{\text{STF}}

\title{Sub-subleading soft gravitons and large diffeomorphisms}

\author[a]{Miguel Campiglia}
\author[b]{Alok Laddha}
\affil[a]{Facultad de Ciencias, Montevideo, Uruguay}
\affil[b]{Chennai Mathematical Institute, Chennai, India}
\date{}

\begin{document}
\maketitle

\thispagestyle{empty}

\let\oldthefootnote\thefootnote
\renewcommand{\thefootnote}{\fnsymbol{footnote}}
\footnotetext{Email: campi@fisica.edu.uy, aladdha@cmi.ac.in}
\let\thefootnote\oldthefootnote

\begin{abstract}
We present strong evidence  that the sub-subleading soft theorem in semi-classical (tree level) gravity discovered by Cachazo and Strominger is equivalent to the conservation of  asymptotic charges associated to a new class of vector fields not contained within the previous extensions of BMS algebra. 
Our analysis crucially 
relies on analyzing the hitherto established equivalences between soft theorems and Ward identities from a  
new perspective.
In this process we  naturally (re)discover a class of  `magnetic' charges at null infinity that are associated to the dual of the Weyl tensor. 
\end{abstract}

\section{Introduction}
The role of BMS group \cite{bms} for quantum gravity in   asymptotically flat spacetimes was extensively studied in the eighties by Ashtekar et.al. \cite{aaprl,as,aajmp,aabook,nut4} (see \cite{aanull} for a recent review). The subject experienced a renaissance recently due to seminal work by  Strominger and collaborators \cite{strom1,stromst} that relates this asymptotic symmetry group with  Weinberg's soft graviton factorization theorem \cite{weinberg}.  The new insight led to further developments in which groups larger than BMS  have emerged as candidate symmetries of quantum gravity. On the scattering amplitude side, Strominger and Cachazo \cite{cs} showed how Weinberg's theorem  can be extended to sub and sub-sub leading order in the soft graviton energy. (For beautiful and alternative derivations of these theorems we refer the reader to \cite{broedel,bern}).
The subleading factorization was identified in \cite{stromvir} with  Ward identities of the `extended' BMS group  of Barnich and Troessaert \cite{barnich}. In \cite{cl1,cl2} we proposed that the subleading relation is best understood in terms of a different extension of  BMS  referred to as `generalized' BMS group.

 Following this line of reasoning, it appears that each factorization theorem 
 is nothing but a Ward identity of certain (spontaneously broken) symmetries of semi-classical gravity. For tree-level quantum gravity amplitudes, three factorization theorems are known so far (and there are good reasons to believe that even at tree level, there may not be anymore \cite{loopcorr}).
As we have an understanding of the symmetries which give rise to the first two of these theorems, 
a natural question to ask is, if the sub-subleading soft theorem is also equivalent to certain Ward identities in tree-level quantum gravity.

Drawing on our previous work regarding symmetries associated to Low's theorem in massless QED \cite{maxwell}, in this paper we provide strong evidence that such a symmetry exists and is generated by vector fields on the conformal sphere at null infinity, which vary linearly along the null generators. The main ideas and results were already presented in \cite{prl}. Here we provide all the details of the analysis that were alluded to in \cite{prl}.  An outline summarizing the conceptual line of thought that we employ here is summarized in the next section.

\section{Outline}
We begin by presenting  by now the well established relationship between Ward identities of the so called generalized BMS group (henceforth denoted by ${\cal G}$) and leading as well as sub-leading soft theorems from a  different perspective. The seminal work of Strominger et. al. \cite{stromst} established the equivalence between Ward identities associated to supertranslation symmetry and Weinberg's soft graviton theorem using charges associated to  supertranslations which were \noindent{{\bf (a)}} derived in Bondi gauge and \noindent{{\bf (b)}} there was one charge associated to each supertranslation generator. At the outset there are two aspects of this equivalence which warrant further investigation. The first one being that the soft theorems are themselves derived in de Donder gauge and hence  we can ask if it is possible to also compute the charges associated to asymptotic symmetries in de Donder gauge. Second and perhaps more serious issue arises from the fact that there are ``$2\times\infty$'' number of soft theorems due to 2 polarizations of the soft gravitons (the infinity stands for the soft momentum direction) whereas there is only one charge for each supertranslation generator.  The first issue is simply a technicality as the asymptotic charges are expected to be gauge invariant and can be derived in any gauge (Bondi or de Donder) that one desires. The second discrepancy was resolved by Strominger \cite{strom1} using an ingenious idea. As far as  perturbative gravitational scattering processes are concerned, there exists a  constraint which relates positive and negative helicity soft gravitons referred to as Christodoulou-Klainerman (CK) condition. This constraint is naturally obeyed by all asymptotically flat geometries which are ``in a neighborhood  of Minkowski space". CK condition implies that the soft theorem for positive helicity soft graviton implies soft theorem for negative helicity graviton and vice versa. Whence there remain $1 \times \infty$ number of independent soft theorems, in accordance with the number of Ward identities.\\

In this work we first revisit these two aspects of the equivalence. Namely, we  show that one can derive the charges associated to  ${\cal G}$ in de Donder gauge, there by placing both, the Ward identities and soft theorems on an equal footing.  We also show that the so-called CK condition can be understood as the vanishing of a particular ``magnetic charge" associated to supertranslations. Thus for each generator of supertranslation there really are two charges. One is the charge derived and used in \cite{stromst} and the other is a magnetic charge which when set equal to zero  is precisely the CK condition.  Our analysis of deriving asymptotic charges in de Donder gauge is predicated upon an understanding of BMS symmetry as residual large gauge transformations (of perturbative gravity) in de Donder gauge. That is, we consider vector fields which satisfy
\begin{equation} \label{waveeqoutline}
\square\xi^{a}\ =\ 0 ,
\end{equation}
and which do not fall off to zero at null infinity. As we demonstrate in section \ref{BMS-dedonder}, all the generators of ${\cal G}$ can be understood as large diffeomorphisms with different asymptotic conditions on various components of $\xi^a$ at null infinity. (This idea first appeared in a paper by Avery and Schwab \cite{avery}). In  section \ref{BMS-dedonder}  we also compute the asymptotic charges associated to these large gauge transformations via covariant phase space methods in perturbative gravity (reviewed in section \ref{sec3}) . We show how the asymptotic charges one computes using this method match precisely the asymptotic charges associated to ${\cal G}$  \cite{as,stromst,stromvir,cl2}. This resolves the first less analyzed aspect of the equivalence we mentioned above.
The reason we go through all this trouble is however  not to merely recycle known results from a different perspective. The main reason is the following: Our goal is to see if  the sub-subleading theorem of perturbative gravity can be also understood as Ward identities associated to certain symmetries. As all the charges corresponding to ${\cal G}$ generators are equivalent to the leading and subleading theorems, it is clear that such a symmetry, if it exists has to be an extension of ${\cal G}$. It turns out such extensions are easier to probe in de Donder gauge. 
A back of the envelope computation indicates that a charge associated to a vector field will correspond to a sub-subleading soft graviton if the  non-trivial sphere vector field component at null infinity $\xi^{A}$  is linear in $u$ (for ${\cal G}$ generators such components are  independent of $u$). Hence we seek solutions to the wave equation (\ref{waveeqoutline}) whose $O(r^{0})$ sphere vector field components are linear in $u$. We will show how there exists a class of such vector fields for which the associated  asymptotic charge, as computed through covariant phase space methods are such that the corresponding Ward identities are precisely equivalent to the sub-subleading soft theorem for a specific combination of positive and negative helicity gravitons. The missing ingredient in proving a complete equivalence between sub-subleading theorem and the asymptotic symmetries is 
that so far we do not have a first principle derivation of the charges whose associated Ward identities are equivalent to the soft theorem for a graviton of orthogonal helicity. Trying to hunt down this ``missing charge" leads us to yet another perspective on the asymptotic charges in terms of electric and magnetic parts of the Weyl tensor. This perspective was already known and investigated by Ashtekar and Sen in \cite{nut4} in which each supertranslation generator yields  two type of charges. The charge obtained by integrating the electric part of the Weyl tensor is the  supermomentum flux which was used as asymptotic charge by Strominger et.al. in \cite{stromst}. The other, lesser known charge --referred to as  NUT supermomentum-- is precisely  the charge obtained from the magnetic part of the Weyl tensor. In section \ref{sec7} we revisit these ideas from a covariant phase space perspective and by considering such magnetic charges for ${\cal G}$ generators and elaborating on their roles in soft theorems, conjecture that the magnetic charges associated to the new symmetries provide the ``missing charge''.

Throughout the paper (except in section \ref{sec7}) we work in the context of gravity coupled with massless scalar field. The massless scalar particles will play the role of external `hard' particles in the soft theorems. The reason for this choice is that computations become simpler. It should however be straightforward to extend the analysis to gravity coupled to other fields or to pure  perturbative gravity. 

\section{Linearized gravity coupled to massless scalar field}\label{sec3}
The system we will be studying in this paper is perturbative gravity coupled to a massless scalar field $\varphi$.   In de Donder gauge, the field equations for the metric perturbation $h_{ab}$  are given by,
\begin{equation} \label{waveeqh}
\begin{array}{lll}
\square \overline{h}_{ab}\ =\ -2 \T_{ab}
\end{array}
\end{equation}
where $\overline{h}_{ab}=h_{ab}\ -\frac{1}{2}\eta^{mn}h_{mn}h_{ab}$,  $\T_{ab}$ the stress tensor of the scalar field and $\square$ the flat space wave operator.  $\overline{h}_{ab}$ satisfies the de Donder gauge condition
\be
\nabla^b \hb_{ab} =0.
\ee
Indices are raised and lowered with the flat metric $\eta_{ab}$ and  $\nabla_a$ is the flat derivative, $\nabla_a \eta_{bc}=0$.

  We will study these equations in retarded  coordinates $(u,r,\hat{x})$ as they are most suitable for massless fields. In these coordinates, we specify ``radiative free data" at future null infinity and solve the equations recursively in $1/r$. The matter free data is given by a function $\phi(u,\xh)$ at null infinity that specifies the  leading $r \to \infty$  term of the  scalar field,
 \be
 \varphi(u,r,\xh) = \frac{\phi(u,\xh)}{r} +O(r^{-2}).
 \ee
  The gravitational free data $C_{AB}(u,\xh)$ is given by the leading angular components of the metric perturbation:
  \be
  h_{AB}(u,r,\xh) = r \, C_{AB}(u,\xh) + \ldots
  \ee
(capital indices denote sphere components).  A solution to the wave equation (\ref{waveeqh}) can then be written as $h_{ab}= h^{(C)}_{ab}+ h^{(\phi)}_{ab}$ where  $h^{(C)}_{ab}$, $h^{(\phi)}_{ab}$ satisfy,
\be
\square \hb^{(C)}_{ab} = 0 , \quad \quad \square \hb^{(\phi)}_{ab} = -2\T_{ab}.
\ee
The metric perturbation $h^{(C)}_{ab}$ is determined by the gravitational radiative  data $C_{AB}$ and the metric perturbation $h^{(\phi)}_{ab}$ is determined by the  radiative matter data $\phi$. The detailed asymptotic form of  $h^{(C)}_{ab}$ and $h^{(\phi)}_{ab}$ are given in appendices \ref{appfreeh} and \ref{appsourcedh} respectively.

\subsection{Asymptotic charges} \label{asymcharges}
The  symplectic potential density of gravity coupled to a scalar field is given by: 
\be \label{theta}
\theta^a(\delta) = \theta^a_{\grav}(\delta) + \theta^{a}_{\matt}(\delta)
\ee
where
\ba
\theta^a_{\grav}(\delta)  & = &\frac{1}{2}\sqrt{g}\left(g^{bc}\delta\Gamma^{a}_{bc}[g] - g^{ab}\delta\Gamma^{c}_{cb}[g]\right) , \label{thetagrav} \\ 
\theta^{a}_{\matt}(\delta) & =& - \sqrt{g} g^{ab} \partial_b \varphi\delta\varphi .
\ea
Given a vector field $\xi^a$, the covariant phase space charge \cite{abr,leewald} at null infinity is determined by the condition:
\be
\delta Q_\xi = \lim_{t \to \infty} \int_{\Sigma_t} dS_a (\delta \theta^a(\delta_\xi) -\delta_\xi \theta^a(\delta)), \label{cpsdelQ}
\ee
where $\Sigma_{t}$ is a $t=$ constant surface that approaches null infinity as $t \to \infty$. It is understood that in this limit the integrand of (\ref{cpsdelQ}) is evaluated by keeping $u=$ constant, as appropriate for massless fields (see \cite{massivebms} for how this changes in the presence of massive fields). For the purposes of making contact with the tree-level soft theorems, we are interested in keeping terms in the charge that are quadratic in the scalar field radiative data $\phi$ (referred to as `hard' part) and linear in the gravitational radiative data $C_{AB}$ (referred to as `soft' part of the charge). 
From the splitting (\ref{theta}) we can write  
\be
Q_\xi= Q^{\grav}_\xi +Q^{\matt}_\xi,
\ee
where each part is defined as in (\ref{cpsdelQ}) with $\theta^a$ replaced by $\theta^a_{\grav}$ or $\theta^a_{\matt}$.
One can verify that the matter contribution is given by:
\be \label{Qmatt}
Q^{\matt}_\xi =- \lim_{t \to \infty} \int_{\Sigma_t} dS_a \sqrt{g} \, \T^{a}_{\ b} \,  \xi^b.
\ee
In the limit and setting of interest, the metric $g_{ab}$ in (\ref{Qmatt}) can be replaced by the flat metric $\eta_{ab}$, and $Q^{\matt}_\xi$ becomes quadratic in the scalar field. It thus contributes to the `hard' charge.

For the limit and setting of interest the gravitational part $Q^{\grav}_\xi$ can be computed by keeping terms that are linear in the metric perturbation $h_{ab}$. That is, 
it suffices to work with the symplectic potential $\theta^{a}_{\lin}$ of linearized gravity. From (\ref{thetagrav}) one finds it is given by (after dropping total variation terms):
\be \label{thetalin}
\theta^a_{\lin} = \frac{\sqrt{\eta}}{2} ( \delta \hb^{bc}  \Gamma^a_{bc} + \frac{1}{2} \delta \hb^{ab} \partial_b \hb   ),
\ee
where  $\Gamma^a_{bc}$ refers now to the linearized Christoffell symbols:
\be
\Gamma^{a}_{bc}\ =\ \frac{1}{2}\eta^{ad} (\nabla_b h_{cd}+ \nabla_c h_{bd}- \nabla_d h_{bc}).
\ee
When condition (\ref{cpsdelQ}) is written for the linearized gravity symplectic potential, the resulting expression is automatically the total variation of a charge given by:
\be \label{Qgrav}
Q^{\grav}_\xi[h] = \lim_{t \to \infty} \frac{1}{2} \int_{\Sigma_t} dS_a \sqrt{\eta} \big(\Gamma^a_{bc} \delta_\xi \hb^{bc}-\delta_\xi \Gamma_{bc}^a \hb^{bc}   + \frac{1}{2}\delta_\xi \hb^{a b} \partial_b \hb -\frac{1}{2}   \hb^{a b} \partial_b \delta_\xi \hb \big).
\ee
The charge is linear in the metric perturbation $h_{ab}$. But from the previous section we have that 
$h_{ab}$ is a sum of two components,
\be
h_{ab} = h^{(C)}_{ab} + h^{(\phi)}_{ab},
\ee
with $h^{(C)}_{ab}$ the `free' metric perturbation that depends (linearly) in the gravitational data $C_{AB}$ and $h^{(\phi)}_{ab}$ the `sourced' metric perturbation that depends (quadratically) on the scalar field data $\phi$. Accordingly, the charge (\ref{Qgrav}) takes the form of a sum:
\be
Q^{\grav}_\xi[h] = Q^{\grav}_\xi[h^{(C)}]  + Q^{\grav}_\xi[h^{(\phi)}] .
\ee
It then follows that $Q^{\grav}_\xi[h^{(C)}]$ yields the \emph{soft} part of the charge, whereas $Q^{\grav}_\xi[h^{(\phi)}]$ contributes to the \emph{hard} part of the  charge. In summary, the total charge can be written as:
\be
Q_\xi = Q^{\hard}_\xi + Q^{\soft}_\xi
\ee
with
\ba
Q^{\soft}_\xi & = & Q^{\grav}_\xi[h^{(C)}] \\
Q^{\hard}_\xi & =  & Q^{\matt}_\xi + Q^{\grav}_\xi[h^{(\phi)}].
\ea
 As we will see in section \ref{BMS-dedonder},  $Q^{\grav}_\xi[h^{(\phi)}]$ is zero for ${\cal G}$ vector fields but will be non-trivial for the symmetries which lead to sub-subleading theorem. 
 
 We conclude the section by introducing notation for later reference. Taking $\Sigma_t$ as a $t=u+r$=constant surface, we write the total charge as
 \be
 Q_\xi = \lim_{t \to \infty} \int du d^2\xh \sqrt{q} \, \rho
 \ee
  where the density being integrated is a sum of five terms:
\be
\rho = \rho_T + \rho_1+\rho_2+ \rho_3+\rho_4, \label{rhohardtotal}
\ee
corresponding to the terms appearing Eqns. (\ref{Qmatt}) and (\ref{Qgrav}), namely:
\ba
\rho_T & = & - r^2 T^t_{\, a} \xi^a  \label{rhoT} \\
\rho_1 &= & \frac{r^2}{2} \Gamma_{ab}^t \delta_\xi h^{ab} \label{rho1} \\
\rho_2 &= & - \frac{r^2}{2} \delta_\xi \Gamma_{ab}^t \hb^{ab} \label{rho2} \\
\rho_3 &= &  \frac{r^2}{4} \delta_\xi \hb^{t b} \partial_b \hb \label{rho3}\\
\rho_4 &= &  \frac{r^2}{2} \hb^{t b} \partial_b( \nabla_c \xi^c  \label{rho4})
\ea
(above we used that $\eta^{bc} \Gamma^a_{bc}=0$ and $\delta_\xi \hb = -2 \nabla_c \xi^c$).

\section{Generalized BMS in de Donder gauge}\label{BMS-dedonder}
In this section we show how the generalized BMS group (which is naturally defined in Bondi gauge) and its associated asymptotic charges can be analyzed from the de Donder gauge perspective. That is, we consider certain generators of residual diffeomorphisms for linearized gravity in de Donder gauge which are \noindent{\bf (a)} asymptotically divergence-free and \noindent{\bf (b)} have the same fall-off behaviour as the ${\cal G}$ generators, and show that the corresponding charges coincide with the known charges associated to supertranslation and $\textrm{Diff}(S^{2})$ vector fields of ${\cal G}$. The analysis of BMS algebra (in arbitrary dimensions)  in de Donder gauge was first given in the seminal work of  \cite{avery}.

We compute the charges associated to such ``large" diffeomorphisms using covariant phase space techniques. We show that these charges contain terms that diverge logarithmically with $r$. \emph{However the fact that the vector fields 
satisfy the wave equation implies that this logarithmically divergent term  vanish}.  The finite part of the charge turns out to be precisely equal to the charges associated to generalized BMS algebra. 

The analysis in this section will  set the stage for exploring a new class of symmetries which give rise to sub-subleading theorem.

In the de Donder gauge, the residual gauge transformations are given by vector field $\xi^{a}$ which satisfy 
\be
\begin{array}{lll} \label{boxxi}
\square\xi^{a}\ =\ 0\\
\end{array}
\ee
and which do not vanish at null infinity.  In order to understand ${\cal G}$ as residual symmetry in de Donder gauge, we consider the following ansatz for $\xi^{a}$ in retarded coordinates $(u,r,\hat{x})$:
\be \label{asymxibms}
\begin{array}{l}
\xi^r  =  r \xioner +O(1) \\
\xi^u  =   \xiou + O(r^{-\epsilon})  \\
\xi^A  =  \xioA + r^{-1}\overset{(-1)}{\xi^A} +O(r^{-1-\epsilon})  
\end{array}
\ee
The leading terms of the wave equation (see appendix \ref{appboxxi}) implies that $\xioner$, $\xiou$ and $\xioA$ are all $u$-indepedendent. Next we impose the defining condition of generalized BMS group $\cal G$, namely that $\xi^{a}$  is asymptotically divergence free \cite{cl1}:
 \be \label{asymdivfree}
 \lim_{r\rightarrow\infty}\nabla_{a}\xi^{a} = 0.
 \ee
 This condition implies that:
 \be \label{div0}
 3 \xioner + \partial_{u} \xiou +\ D_{A}\xioA = 0.
 \ee
On the other hand, the vanishing of $\square \xi^u$ at order $r^{-1}$ yields (see appendix \ref{appboxxi}):
\be \label{boxxium1}
- \partial_u \xiou +  \xioner +D_A \xioA=0.
\ee
From equations (\ref{div0}) and (\ref{boxxium1}) one finds:
\be \label{gralbmsvfcond}
\partial_u \xiou = \frac{1}{2} D_A \xioA, \quad \quad \xioner= - \frac{1}{2} D_A \xioA .
\ee
The first condition yields 
\be
\xiou = \frac{u}{2} D_A \xioneA + f(\xh),
\ee
with $f$ an arbitrary function on the sphere that appears as an integration `constant'. Calling
\be
\xioA= V^A, \quad  \alpha = \frac{1}{2}D_A V^A,
\ee
the  vector field takes the form:
\be \label{gralbmsvf}
\xi^a = V^A \partial_A + (u \alpha +f) \partial_u - \alpha r \partial_r + \ldots,
\ee
where the dots represent subleading terms as in (\ref{asymxibms}). This is precisely the form of the generalized BMS vector field given in \cite{cl1}. Setting $V^A=0$ one obtains a supertranslation vector field and setting $f=0$ one obtains the sphere vector fields associated to subleading soft graviton theorem.  

We now proceed to compute the associated charges along the lines presented in the previous section. It will be convenient to discuss separately the supertranslation and sphere vector field cases, particularly as they require different $u \to \infty$ fall-offs on the radiative data.

\subsection{Supertranslation charges} \label{STcharge}
For supertranslation charges one can use standard radiative phase space fall-offs at $u \to \pm \infty$ \cite{as}:
\be \label{falloffST}
C_{AB}(u,\xh) = C^\pm_{AB}(\xh)+O(|u|^{-\epsilon}), \quad \quad \phi(u,\xh) = O(|u|^{-\epsilon}).
\ee
We now consider the general charge formulae given in section \ref{asymcharges} for the case of a supertranslation vector field
\be
\xi^a_f = f(\xh) \partial_u + \ldots
\ee
where $\xi^r_f=O(1)$ and $\xi^A_f = O(r^{-1})$.  Given the $r \to \infty$ fall-offs described in the appendices, one finds the densities (\ref{rhoT}) to (\ref{rho4}),  in the $r \to \infty$ limit, are given by:
\be
\rho_T  =   \xiou \Ttwo_{uu} =  f \dot{\phi}^2 
\ee
\be
\rho_2 =\rho_3 = \rho_4 =0.
\ee
\be
\rho_1 = \overset{(-1)}{\Gamma^t_{rA}} \overset{(-1)}{\delta_\xi h^{r A}}+\frac{1}{2} \overset{(1)}{\Gamma^t_{AB}}  \overset{(-3)}{\delta_\xi h^{AB}} = \frac{1}{2} \partial_u\big( C_{AB} D^A \overset{(-1)}{\xi^B} \big),
\ee
where in the last equality we dropped a total sphere divergence. The component $\overset{(-1)}{\xi^A}$  is determined by the preservation of the metric perturbation fall-offs. Specifically for a supertranslation vector field one finds:
\be
\delta_\xi h_{Ar}= O(r^{-1}) \iff \overset{(-1)}{\xi^A} = -D^A f.
\ee
Thus the total charge is given by:
\be
Q_{\xi_f}= \int du d^2\xh \sqrt{q} ( f \phid^2 - \frac{1}{2} D^A D^B f \partial_u C_{AB}),
\ee
which  corresponds to the well known expression of supertranslation charge \cite{as,stromst}.
\subsection{Sphere vector field charges} \label{sphvfcharge}
For the sphere vector fields 
\be \label{sphvf}
\xi^a_V = V^A \partial_A + u \alpha  \partial_u - \alpha r \partial_r + \ldots,
\ee
the charges are defined on a subspace of radiative data  where $C_{AB}$ satisfies the stronger fall-offs \cite{cl2}:
\be \label{falloffvf}
C_{AB}(u,\xh) = O(|u|^{-1-\epsilon}),
\ee
(for $\phi$ we keep the same fall offs as before). To simplify the analysis, we discuss separately the `hard' and `soft' part of the charge.
\subsubsection{Hard part} \label{hardpartsphvfsec}
From the $r \to \infty$ fall-offs described in the appendices, one finds that for a vector field (\ref{asymxibms}) there is a  potential $\log r$ divergence in the hard charge density. The divergent term arises in the piece $\rho_1$ (\ref{rho1}) and is given by:
\ba
\overset{(\ln)}{\rho_1} &= &  \overset{(-2 \ln)}{\Gamma^t_{ru}}  \overset{(0)}{\delta_\xi h^{ru}} + \frac{1}{2} \overset{(\ln)}{\Gamma^t_{AB}}  \overset{(-2)}{\delta_\xi h^{AB}} \\
& =&  \frac{1}{2}\mu(\xioner - \partial_u \xiou + D_A \xioA) \label{logvf}.
\ea
However, this term  vanishes by virtue of the wave equation, Eq. (\ref{boxxium1}). Whence  it turns out that the vector fields satisfying wave equation and having fall-offs given in Eq.(\ref{asymxibms}) yield finite charges. 

The remaining contribution to the hard charge are finite. From the fall-offs described in the appendices, one finds the following  limiting expressions for $\rho_T$ and (hard part of)  (\ref{rho1}) to (\ref{rho4}):
\be
\rho_T =  \xiou \Ttwo_{uu}  + \xioA\Ttwo_{u A} = \phid ( \xioA \partial_A \phi +\xiou \phid),
\ee
\be
\rho_1 =  \overset{-2 }{\Gamma^t_{ru}}  \overset{0}{\delta_\xi h^{ru}} + \frac{1}{2} \overset{0}{\Gamma^t_{AB}}  \overset{-2}{\delta_\xi h^{AB}} =0
\ee
\be
\rho_3= \frac{1}{4} \overset{(0)}{\delta_\xi \hb^{t u}}  \overset{-2 }{\partial_u \hb} =0
\ee
and
\be
\rho_2 = \rho_4=0.
\ee
The vanishing of $\rho_1$ and $\rho_3$ is due to the fact that for generalized BMS vector field $\overset{0}{\delta_\xi h^{ru}}=\overset{(0)}{\delta_\xi \hb^{t u}} =0$ and $q_{AB}\overset{-2}{\delta_\xi h^{AB}}=0$, together with the fact that $\overset{0}{\Gamma^t_{AB}} \propto q_{AB}$ (see appendices). Thus, as anticipated in the previous section, there is no hard contribution from the gravitational part of the charge. The hard charge is then given by:
\be \label{rhohardV}
\rho^{\hard}_V= \rho_T =  \phid( V^A \partial_A + u \alpha \phid),
\ee
which represents the scalar field contribution to the hard charge computed in \cite{cl2} for pure gravity.
\subsubsection{Soft part}
For the soft part one finds
\be
\rho_3=\rho_4=0,
\ee
\be
\rho_2 =-\frac{1}{2} \overset{(1)}{\delta_\xi \Gamma_{AB}^t} \overset{(-3)}{h^{AB}}=-\frac{1}{2} C_{AB} D^A D^B \xioner,
\ee
and
\be
\rho_1 = r  \overset{(1)}{\rho_1} + \overset{(0)}{\rho_1}
\ee
\be
\overset{(1)}{\rho_1}=\frac{1}{2} \overset{(1)}{\Gamma^t_{AB}}  \overset{(-2)}{\delta_\xi h^{AB}}= \frac{1}{2} \partial_u C_{AB} D^A \overset{(0)}{\xi^B}
\ee
\ba
\overset{(0)}{\rho_1}& =& \overset{(-1)}{\Gamma^t_{rA}} \overset{(-1)}{\delta_\xi h^{r A}}+\frac{1}{2} \overset{(1)}{\Gamma^t_{AB}}  \overset{(-3)}{\delta_\xi h^{AB}} +\frac{1}{2} \overset{(0)}{\Gamma^t_{AB}}  \overset{(-2)}{\delta_\xi h^{AB}} \\
&=& \frac{1}{2} \partial_u\big( C_{AB} D^A \overset{(-1)}{\xi^B} \big)-\frac{1}{2}C_{AB}D^A D^B \xioner + \frac{1}{2} \overset{(0)}{\partial_u h_{AB}} D^A \xioB
\label{finalrho11vf}
\ea
(in the last line we discarded a total sphere divergence). To obtain the charge we  write $r=t-u$ and take $t \to \infty$ with $u$ fixed. The $O(t)$ term given by $ \overset{(1)}{\rho_1}$ integrates to zero by virtue of the fall-offs (\ref{falloffvf}). These fall-offs also imply  the part proportional to $\overset{(-1)}{\xi^A}$ in (\ref{finalrho11vf}) integrates to zero. 
Thus, the total soft charge is given by:
\ba
\rho^{\soft} &= &- u \overset{(1)}{\rho_1} + \overset{(0)}{\rho_1}+ \overset{(0)}{\rho_2} \\
 &= &  \frac{1}{2}C_{AB}(D^A \xioB -2 D^A D^B \xioner) + \frac{1}{2}\overset{(0)}{\partial_u h_{AB}} D^A \xioB
\ea
where in the last line we discarded a total $u$-derivative terms by virtue of  (\ref{falloffvf}).  Writing  the vector field as in section \ref{BMS-dedonder}:
\be
\xioA= V^A, \quad \xioner= -\frac{1}{2}D_A V^A ,
\ee
and  using the expression of $\overset{(0)}{h_{AB}}$ in terms of $C_{AB}$ (Eq. (\ref{freehC})),  we arrive at (discarding total sphere divergences)
\ba
\rho^{\soft}_V & =& \frac{1}{2}C_{AB}(2D^A V^B + D^A D^B D_C V^C- \frac{1}{2} \Delta D^A V^B)  \\
&=&\frac{1}{2}C_{AB}(\frac{1}{2}D^A V^B+ D^A D^B D_C V^C- \frac{1}{2} D^A \Delta  V^B) \label{softvffinal}
\ea
where in the last equality we used the identity: $\Delta D^{(A}V^{B)} = 3 D^{(A} V^{B)} + D^{(A} \Delta V^{B)}$.
Expression (\ref{softvffinal})  precisely coincides with the soft charge computed in  \cite{cl2} in Bondi gauge. When written in stereographic coordinates $(z,\zb)$,  it takes the form of the soft charge first presented in \cite{stromvir}:
\be
\rho^{\soft}_V =  \frac{1}{2}C^{zz} D^3_z V^z + c.c.
\ee

\section{Extracting charges from the sub-subleading theorem} \label{sec5}
The sub-subleading soft theorem takes the form
\be \label{ssl}
[\lim_{\w \to 0} \omega^{-1} {\cal M}_{n+1}(\omega\hat{q},k_{1},\dots,k_{n})]_{\text{finite}}\ =\ S^{(2)}{\cal M}_{n}(k_{1},\dots,k_{n})
\ee
where we are discarding divergent $O(\w^{-2})$ and $O(\w^{-1})$ terms,  keeping only the finite piece. The factor  $S^{(2)}$ is described below. Since in Fourier space diving by frequency amounts to an integral over time $u$,
\be
\omega^{-1} \tilde{F}(\omega)\ = -i \int_{-\infty}^{\infty} du e^{i\omega u} \int_{-\infty}^{u}F(u') du',
\ee
we are motivated to define the prospective soft charge corresponding to sub-subleading theorem as follows. Given a symmetric, trace-free sphere tensor $Y^{AB}$ let us define the soft charge as \cite{prl}:
\be
\Qs_{Y}:= \int_{-\infty}^{\infty} du \int_{-\infty}^u du' \int d^2 w \sqrt{\gamma} \; Y^{ww} D_w^4 C^{ww}(u',\qh)  + c.c. ,\label{QsT1}
\ee
where $(w,\wb)$ are stereographic coordinates for $\qh$ and $ \sqrt{\gamma} = 2/(1+ w \wb)^2$ the area element. To simplify the discussion we now take $Y^{\wb \wb}=0$ and discuss the general case towards the end of this section.

Recalling that $C_{\wb \wb}(\w,\qh)=\frac{\sqrt{\gamma}}{2 \pi i} a_-(\w,\qh)$ and taking into account the tensorial structure of sphere derivatives, the proposed charge can be written as:
\be
\Qs_Y = -\lim_{\w \to 0} \w^{-1} \frac{1}{2 \pi} \int d^2 w \, Y^{w w}( \partial^4_w +\frac{2 \wb}{1+w \wb} \partial^3_w) a_-(\w,\qh) . \label{QsT2}
\ee
The insertion of this operator in a scattering amplitude can be evaluated with the sub-subleading soft theorem:
\be
[\lim_{\w \to 0} \w^{-1} \bra \out | a_\pm(\w,\qh)  \S |\in \ket]_{\rm finite}=  \sum_i \Stwo_\pm(q,p_i) \bra \out | \S | \in \ket, \label{ssst}
\ee
where the sub-subleading soft factor is the second order differential operator,
\be
\Stwo_\pm(q,p) = (2\,  p \cdot q)^{-1}(\e^\mu_\pm q^\nu J_{\mu \nu})^2 .\label{S2}
\ee
One can check that when $(\partial^4_w +\frac{2 \wb}{1+w \wb} \partial^3_w) $ acts on the soft factor  (\ref{S2}) the result is proportional to Dirac deltas and its derivatives. The sphere integral can then be evaluated, resulting in:
\begin{multline}
\frac{1}{2 \pi} \int d^2 w \, Y^{w w}( \partial^4_w +\frac{2 \wb}{1+w \wb} \partial^3_w) \Stwo_-(q,p) = \\ - \frac{1}{2} E  D^2_z Y^{z z} \partial_E^2 + 2 D_z Y^{z z} \partial_E \partial_z - 3 E^{-1} Y^{z z} D_z \partial_z -2 E^{-1} D_z Y^{zz} \partial_z =:  \Y_Y \label{YY}
\end{multline}

As in \cite{stromlow}, in order to interpret the soft theorem as a Ward identity we now seek for a hard charge $\Qh_Y$ that generates the action (\ref{YY}) via Poisson brackets:
\be
\{ b, \Qh_Y\}  = i \Y_Y b,  \label{PBbQhY}
\ee
where  $b$ is the mode function of the external (scalar) hard particles,
\be
b(E,\xh) = 4 \pi i \int_{-\infty}^\infty du \, e^{i E u} \phi(u,\xh).
\ee
In terms of the mode functions, the symplectic product of the scalar field reads:
\be
\O(\delta,\delta')= \frac{2 i}{(4 \pi)^2}\int d^2 V \int_{-\infty}^\infty \frac{dE}{2 \pi} E  \, \delta b \ \delta' b^*. \label{sp}
\ee
Since $ i \Y_Y b$  is homogenous in $b$, the candidate charge can be computed by: 
\be
\Qh_Y = \frac{1}{2}\O(i \Y_Y b,b) . \label{QhT1}
\ee
There are three types of terms appearing in (\ref{QhT1}) of the form:
\ba
(4 \pi)^{-2}\int_{-\infty}^\infty \frac{d E}{2\pi}  \, A b(E) \, b^*(E) = - \int_{-\infty}^\infty du A\phi(u) \, \phi(u)  \\
(4 \pi)^{-2}\int_{-\infty}^\infty \frac{d E}{2\pi}  \, A E \partial_E b(E) b^*(E) = - \int_{-\infty}^\infty du \, u A\phi(u) \partial_u \phi(u) \\
(4 \pi)^{-2}\int_{-\infty}^\infty \frac{d E}{2\pi}  \, A E^2 \partial^2_E b(E) b^*(E) = -\int_{-\infty}^\infty du \, u^2 A\phi(u) \partial^2_u \phi(u) 
 \ea
where $A$ denotes a sphere differential operator. Using these identities, one finds, after some integration by parts:
\begin{multline}
\Qh_Y = - \int d^2 V \int du \big( 3 Y^{zz} \partial_z \phi  \partial_z \phi  +2 u \, D_z Y^{zz} \partial_z \phi \, \phid \\
 +  D^2_z Y^{zz}(- \phi^2 + \frac{u^2}{2} \phid^2) \big). \label{QY}
\end{multline}
One can then explicitly check that the Poisson bracket between $b$ and $\Qh_Y$ satisfies (\ref{PBbQhY}) as desired. \\

It is straightforward to extend the previous analysis to the  case of a general real $Y^{AB}$. The associated hard charge is then given by expression (\ref{QY}) plus its complex conjugate.   In covariant notation, it takes the form:
\begin{multline} \label{finalQY}
\Qh_Y = - \int d^3 V  \big( 3 Y^{AB} \partial_A \phi  \partial_B \phi +2 u \, D_A Y^{AB} \partial_B \phi \, \phid\\
  +D_A D_B Y^{AB}(- \phi^2  + \frac{u^2}{2} \phid^2) \big)
\end{multline}

By the standard reasoning (see e.g. \cite{massivebms,stromlow}) one concludes that the sub-subleading soft theorem (\ref{ssl}) implies the S-matrix commutes with the charge
\be
\Q_Y = \Qh_Y + \Qs_Y,
\ee
with $\Qh_Y$ given by Eq. (\ref{finalQY}) and $\Qs_Y$ given by Eq. (\ref{QsT1}). Conversely, one can read-off from  $S^{(2)}(q,p)$ the tensor $Y^{AB}$ associated to a positive or negative soft graviton insertion. For a negative helicity soft graviton with direction $(z_s,\zb_s)$ this is given by:
\be \label{YK}
Y^{ww}= \frac{1}{6} \frac{1+ w \wb}{1+z_s \zb_s} \frac{(w-z_s)^3}{\wb-\zb_s} , \quad Y^{\wb \wb}=0,
\ee
which satisfies\footnote{The tensorial structure now is such that $D^4_w$ acts as the `integrated by parts' version of the differential operator in Eq. (\ref{QsT2}), namely:  $D^4_wY^{ww}= \partial^4_wY^{ww} -  \partial^3_w (\frac{2 \wb}{1+w \wb}Y^{ww})$.}
\be \label{D4delta}
\frac{1}{2 \pi} D^4_w Y^{ww} = \delta^{(2)}(w-z_s).
\ee
From relation (\ref{D4delta}) one can show that the Ward identity $\bra \out |[\Q_Y,S] \in \ket=0$ associated to the  tensor (\ref{YK}) reproduces the (negative helicity) sub-subleading relation (\ref{ssl}). Choosing the complex conjugate of (\ref{YK})  leads to the positive helicity soft theorem.\\

We will later  identify the tensor $Y^{AB}$ with a vector field $X^A$ by:
\be
Y^{AB}= (D^{(A} X^{B)})^{\tf}.
\ee
The following identities will then be useful:
\be
D^2_z Y^{zz} + c.c. = D_A D_B Y^{AB} = D \cdot X+ \frac{1}{2} \Delta \,  D \cdot X \label{D2Y}
\ee
and
\be
D_z Y^{zz}\partial_z +cc = D_B Y^{AB} \partial_A = \frac{1}{2}(\Delta X^A +X^A) \partial_A.
\ee

\section{Looking for new symmetries in de Donder gauge} \label{sec6}
As shown in the previous section, the charges in perturbative gravity whose Ward Identities can be derived from the sub-subleading soft theorem are parametrized by symmetric, trace-free tensor fields $Y^{AB}$ on the conformal sphere. This may tempt us to associate these charges to certain generalized symmetries arising perhaps from asymptotically Killing tensor fields. This line of reasoning, while certainly intriguing is made complicated by the fact that there is no natural method to compute charges associated to asymptotic Killing tensors in field theory. (However there is a possibility that by carefully analyzing and extending the methods developed by \cite{jezierski}, one may be able to derive such charges.)  There is a natural analogue of this conundrum in QED. In that case, working  backwards from Low's sub-leading theorem, one obtains asymptotic charges parametrized by vector fields on the sphere \cite{stromlow}. However, as shown in \cite{maxwell}, these charges could be derived from first principle by parametrizing them by $u$-dependent large gauge transformations. Inspired by this,  we will now like to attempt something analogous in the current scenario.

That is, we would like to find \emph{vector fields} whose asymptotic charges reproduce the charges obtained in the last section. As discussed briefly in the outline section, given the form of both hard and soft parts, one is lead to conclude that the vector fields should have an extra power of $u$ with respect to generalized BMS vector fields $\xi_V$,  or two extra powers of $u$ with respect to supertranslations vector fields $\xi_f$ (or both). 
Requiring that the vector field satisfies the wave equation, one is lead to an ansatz of the form (see appendix \ref{appboxxi}):
\ba
\xi^r & = & r^2 \xitwor + r \xioner +O(r^0) \nonumber \\
\xi^u & = & r^2 \xitwou + r \xioneu + \xiou + O(r^{-\epsilon})  \label{newsymxi} \\
\xi^A & = & r \xioneA +  \xioA + O(r^{-1}) . \nonumber 
\ea
The wave equation implies the leading terms $\xitwor $, $\xitwou$ and $\xioneA$ are $u$-independent. These leading terms play the role of `free data' in terms of which subleading terms are determined by solving the wave equation (see Appendix \ref{appboxxi}).

In general such type of vector fields will lead to  divergent charges. As we will see, the charges will  have a $t \to \infty$ expansion of the form:
\be  \label{Qexpt}
Q_\xi =t^2 \ln t \overset{(2 \ln)}{Q}_\xi+  t^2 \, \overset{(2)}{Q}_\xi + t \ln t \, \overset{(1 \ln)}{Q}_\xi + t \, \overset{(1)}{Q}_\xi +  \ln t \, \overset{(\ln)}{Q}_\xi +  \overset{(0)}{Q}_\xi +O(t^{-\epsilon}).
\ee
In order to have meaningful finite charges, we need to add counterterms to subtract the divergent terms. Such a procedure is necessarily ambiguous. However in our case such a 
 ``counterterm subtraction" prescription is rendered unambiguous due to the nature of divergent terms. As we will see below,
  the divergent terms turn out to have definite physical interpretation:
\be \label{prescription}
\begin{array}{c}
 \overset{(2)}{Q}_\xi \propto \text{ Supertranslation charge}, \quad  \quad  \overset{(1)}{Q}_\xi \propto  \text{ Diff $(S^2)$ charge} \\
 \\
\overset{(2 \ln)}{Q}_\xi= \overset{(1 \ln) }{Q}_\xi = \overset{(\ln)}{Q}_\xi =0
\end{array}
\ee
The first two conditions are interpreted as subtracting terms due to leading and subleading soft gravitons, and it is inspired by an analogous procedure in the case of subleading soft photon charges in QED \cite{maxwell}.  The terms with logarithms  have a time dependance that is not related to such soft gravitons. We  thus require them to vanish. 
They thus translate into restrictions on the vector field (\ref{newsymxi}). We will find these restricts three of the four independent data in (\ref{newsymxi}). The resulting vector field will be found to be given by:
\be \label{newsymxi2}
\xi^a = r X^A \partial_A+ \ldots , \quad D_A X^A=0,
\ee
where the dots indicate subleading terms that are determined by the `free data' $X^A$ by solving the wave equation $\square \xi^a=0$. Notice that since $X^A$ is restricted to be divergence-free,  the  free data counts as one function on the sphere. Whence just like in the case of supertranslation versus Weinberg soft theorem, we will have Ward identities associated to symmetries that are parametrized by one function on one hand and  two sub-subleading theorem associated to positive and negative helicities respectively on the other.  We will return to this point at the end of the section. 

In the following we compute the divergent and finite contributions to the charges. We focus on the $r \to \infty$ expansion of the charge density (\ref{rhohardtotal}) which will take the form:
\be
\rho =  r^2 \ln r \, \overset{2 \ln}{\rho}+ r^2 \, \overset{2}{\rho}+ r \log r \rhorlr + \log r \rholr+ r \rhor  + \rhoo +O(r^{-\epsilon}).
\ee
Setting  $r=t-u$ yields then the desired $t \to \infty$ expansion.

We start by looking at the logarithmically divergent terms. The condition that they vanish will  yield the form of the vector field (\ref{newsymxi2}). We will then check that polynomially divergent terms satisfy the condition (\ref{prescription}), and finally study the finite charges. To simplify expressions, the dependance on the scalar field will be parametrized in terms of the following quantities:
\be
\mu:= \sint^u \dot{\phi}^2, \quad  \phi^2 , \quad  \Tthree_{uu}, \quad T_{uA}:=\Ttwo_{uA} , \quad T_{AB} :=(\Ttwo_{AB})^{\tf}
\ee
where for the benefit of the reader we recall that $\phi$ is the free data for massless scalar field at null infinity. 

\subsection{Log  divergent terms}
It is clear that due to the power law falls off of $\varphi$ and $\xi^{a}$ with $r$, only the gravitational hard part contains logarithmically divergent terms.  Given the general expression (\ref{rhohardtotal}) and the fall-offs described in the appendices, one finds that the most divergent term associated to the vector field (\ref{newsymxi}) is proportional to $r^2 \ln r$:
\be
\overset{(2 \ln)}{\rho} = \frac{1}{2} \overset{(-1 \ln)}{\Gamma^t_{uu}} \overset{(1)}{\delta_\xi h^{uu}} 
= \frac{1}{2} \phid^2 \, \xitwou.
\ee
Thus, demanding this term to vanish imposes
\be \label{xi2ueq0}
\xitwou= 0.
\ee
From now on we restrict attention to vector fields satisfying this condition.
\subsubsection{$r \log r$}
The term proportional to $r \ln r$ in (\ref{rhohardtotal}) are:
\ba
\overset{(1 \ln)}{\rho_1} &= & \overset{(-2 \ln)}{\Gamma^t_{ru}}  \overset{(1)}{\delta_\xi h^{ru}} + \frac{1}{2}\overset{(-1 \ln)}{\Gamma^t_{uu}}  \overset{(0)}{\delta_\xi h^{uu}} + \overset{(-1 \ln)}{\Gamma^t_{u A}}  \overset{(0)}{\delta_\xi h^{u A}} + \frac{1}{2} \overset{(\ln)}{\Gamma^t_{AB}}  \overset{(-1)}{\delta_\xi h^{AB}}\\
\overset{(1 \ln)}{\rho_2}  &= & -\frac{1}{2} \overset{(0)}{\delta_\xi \Gamma_{rr}^t} \overset{(-1 \ln)}{\hb^{rr} } \\
\overset{(1 \ln)}{\rho_3}  &= &  \frac{1}{4} \overset{(1)}{\delta_\xi \hb^{t u} } \, \overset{(-2 \ln)}{\partial_u\hb} \\
\overset{(1 \ln)}{\rho_4}  &= &  \frac{1}{2} \overset{(-1 \ln)}{\hb^{t r}} \overset{(1)}{\nabla_a \xi^a} 
\ea
Using the expressions from the appendices we get (in what follows we drop total sphere divergences)
\ba
\overset{(1 \ln)}{\rho_1} &= & - \frac{1}{2} \mu \, \partial_u \xioneu + \frac{1}{2} \phid^2 \, \xioneu \\
\overset{(1 \ln)}{\rho_2} &= & - \mu \xitwor \\ 
\overset{(1 \ln)}{\rho_3} &= & 0\\ 
\overset{(1 \ln)}{\rho_4}  &= &  \frac{1}{2} \mu  (4 \xitwor+ \partial_u \xioneu + D_A \xioneA ) 
\ea
Bringing all terms together, one finds:
\be \label{rho1ln}
\overset{(1 \ln)}{\rho}=  \mu ( \xitwor+ \frac{1}{2} D_A \xioneA) + \frac{1}{2} \phid^2 \xioneu
\ee
Since $\partial_u \mu = \phid^2$, one may be tempted to further simplify the expression for the corresponding charge by integrations by parts in $u$.   This however introduces a boundary term since $\sint_{-\infty}^{+\infty} \phid^2 du \neq 0$. Thus, in order for (\ref{rho1ln}) to vanish we need each term to  vanish separately,
\be  \label{logcond}
 \xitwor+ \frac{1}{2} D_A \xioneA=0, \quad \quad \xioneu=0 .
\ee
Combining (\ref{xi2ueq0}), (\ref{logcond}) with $\square \xi^a=0$ one finds the vector field takes the form  (see appendix \ref{appboxxi})
\be  \label{newsymxi3}
\xi^a = (r X^A + \frac{u}{4}(\Delta+5) X^A) \partial_A  + O(r^{-\epsilon})
\ee  
 with $X^A(\xh)$ satisfying $D_A X^A=0$  playing the role of `free data'.   The vector field (\ref{newsymxi3}) will be the candidate vector field associated to sub-subleading charges. Below we show that the associated divergent and finite pieces satisfy the requirements (\ref{prescription}).

\subsubsection{$ \log r$}
Repeating a similar analysis as in the $r \ln r$ case, the terms proportional to $\ln r$ for a vector field with leading components as in (\ref{newsymxi3}) are: 
\be \label{rhologX}
\overset{(\ln)}{\rho_1} =  
 \overset{(-2 \ln)}{\Gamma^t_{ru}}  \overset{(0)}{\delta_\xi h^{ru}} +\overset{(-2 \ln)}{\Gamma^t_{r A}}  \overset{(0)}{\delta_\xi h^{r A}} + \overset{(-2 \ln)}{\Gamma^t_{u A}}  \overset{(0)}{\delta_\xi h^{u A}} + \frac{1}{2} \overset{(\ln)}{\Gamma^t_{AB}}  \overset{(-2)}{\delta_\xi h^{AB}}+\frac{1}{2} \overset{(-1 \ln)}{\Gamma^t_{AB}}  \overset{(-1)}{\delta_\xi h^{AB}} 
\ee
and 
\be
\overset{(\ln)}{\rho_2}   = \overset{(\ln)}{\rho_3}  = \overset{(\ln)}{\rho_4}  =  0
\ee
Using the expressions from the appendices, one finds that all terms in (\ref{rhologX}) are actually zero (up to total sphere divergences) due to $D_A X^A=0$.  Thus, there are no logarithmic divergences associated to the vector field (\ref{newsymxi3}).
\subsection{Polynomially divergent hard terms}
For the vector field (\ref{newsymxi3}) one finds there are no $O(r^2)$ terms in the charge.  For the $O(r)$ piece the only contribution comes from  the stress tensor part and is given by:
\be \label{rho1hardX}
\overset{(1)}{\rho}_T =  \Ttwo_{u A} X^A.
\ee
This has precisely the form of a ${\rm Diff}(S^2)$ hard charge $\rho^{\hard}_X$ (\ref{rhohardV}) (recall $D_A X^A$=0), thus satisfying the required condition (\ref{prescription}).
\subsection{Finite hard charge}
We finally come to the finite part of the hard charge. Here one finds contributions from the stress tensor (\ref{rhoT}) and from the $\rho_1$ term of the gravitational charge (\ref{rho1}). It will be convenient to express the charges in terms of
\ba
T_{uA} & := & \Ttwo_{uA}= \phid \,  \partial_A \phi \\
T_{AB} & := & (\Ttwo_{AB})^{\tf} = (\partial_A \phi \partial_B \phi)^{\tf}.
\ea
and of
\be \label{Yitoxione}
Y^{AB}:= (D^{(A} X^{B)})^{\tf}.
\ee
We start with the stress tensor part. It has a $r \to \infty$ expansion of the form 
\be
\rho_T = r \overset{(1)}{\rho_T} +  \overset{(0)}{\rho_T}.
\ee
 Since the charges are defined by the limit $t \to \infty$ with $u=t-r=$ constant, this gives a finite contribution of the form (see appendix \ref{stressapp}):
\be
\rho_T^{\rm finite} = -u \overset{(1)}{\rho_T} +  \overset{(0)}{\rho_T}=  \Ttwo_{uA}(\overset{0}{\xi^A}-u \,  \overset{1}{\xi^A}) + \Tthree_{u A} \overset{1}{\xi^A}.
\ee
 Up to total  derivatives the last term can be evaluated as (see appendix \ref{appT3uA}):
\be
\Tthree_{u A} \overset{1}{\xi^A} =  D^{(A}\overset{1}{\xi^{B)}} (\partial_A \phi \partial_B \phi)^{\tf}   \label{T3uA}
\ee
and so we write $\rho_T^{\rm finite}$ as
\be \label{rhoTX}
\rho_T^{\rm finite}= (\overset{0}{\xi^A}-u \, \overset{1}{\xi^A})T_{uA} + Y^{AB} T_{AB}.
\ee
For $\rho_1$ we have:
\be
\rho_1 = \overset{-2}{\Gamma^t_{rA}} \,  \delta_\xi \overset{0}{h^{rA}} +\overset{-2}{\Gamma^t_{uA}} \,  \delta_\xi \overset{0}{h^{uA}}+
 \frac{1}{2}  \overset{0}{\Gamma^t_{AB}} \,  \delta_\xi \overset{-2}{h^{AB}} + \frac{1}{2}  \overset{-1}{\Gamma^t_{AB}} \,  \delta_\xi \overset{-1}{h^{AB}}
\ee
From the expressions of the sourced metric perturbation given in Appendix \ref{appsourcedh} one finds:
\be \label{gammasXhard}
\begin{array}{c}
\overset{-2}{\Gamma^t_{rA}}  =  - \sint^u T_{uA} + D_A(\ldots) , \quad  \overset{-2}{\Gamma^t_{rA}} = D_A( \ldots)\\
\overset{-1}{\Gamma^t_{AB}} = q_{AB}( \ldots ) \\
\overset{-1}{\Gamma^t_{AB}} = -\frac{1}{4} T_{AB} + \frac{1}{2}\int^u  D_{(A} T_{B) u} + D_A D_B( \ldots ) + q_{AB}(\ldots ).
\end{array}
\ee
Due to the divergence free property of $\xioneA$ and $\xioA$, only the terms explicitly shown in (\ref{gammasXhard}) give nonzero contribution (up to total sphere divergences). One then finds (after some integration by parts in $u$ and in the sphere):
\be
\rho_1= (u \, \overset{1}{\xi^A}- \overset{0}{\xi^A}) T_{uA} - \frac{1}{4} Y^{AB} T_{AB} +  \frac{u}{2}D_A Y^{AB} T_{u B}.
\ee
Combining this term with the stress tensor contribution (\ref{rhoTX}) one finds the total charge is given by:
\be \label{finalrhohardX}
\rho^\hard=(\rhot)^{\rm finite} + \rho_1 = \frac{3}{4} Y^{AB} T_{AB} +  \frac{u}{2}D_A Y^{AB} T_{u B}.
\ee
Comparing with (\ref{finalQY}) and noting that $D_A D_B Y^{AB}=0$ for divergence-free $X^A$, we see that (\ref{finalrhohardX}) reproduces   (-1/4 times) the charge (\ref{finalQY}) obtained from the soft theorem.

\subsection{Soft charge}
We now compute the soft charge. In the notation of Eq. (\ref{rhohardtotal}) it is a sum of four terms, Eqns. (\ref{rho1}) to (\ref{rho4}), with $h_{ab}$ the `free' linearized metric perturbation associated to $C_{AB}$. The last two terms however do not contribute:  The third one vanishes because $h=0$ and the fourth was already discarded since the vector field was found to be spacetime divergence-free. Thus, only the first two terms contribute:
\be
\rho^\soft= \rho_1+ \rho_2.
\ee
We will find an $r \to \infty$ expansion of the charge density as:
\be
\rho^\soft = r^2 \overset{(2)}{\rho} + r \overset{(1)}{\rho}+  \overset{(0)}{\rho} +O(r^{-\epsilon}),
\ee
which, upon setting $r=t-u$ yields the expansion in $t$:
\be
\rho^\soft = t^2\, \overset{(2)}{\rho} + t \big(-2 u \overset{(2)}{\rho}+ \overset{(1)}{\rho} \, \big)+ \big( u^2  \overset{(2)}{\rho} -u \overset{(1)}{\rho}+ \overset{(0)}{\rho} \, \big)+O(t^{-\epsilon})
\ee

As we will see, in order for the finite charge to be well defined we will need to restrict attention 
to  $C_{AB}$ satisfying
\be \label{strongfallC}
C_{AB}(u,\xh) = O(|u|^{-2 - \epsilon})
\ee
as $u  \to \pm \infty $. It will also be convenient to express the charge density in terms of:
\be
\C_{AB}(u,\xh) := \int^u_{-\infty} C_{AB}(u',\xh) du'.
\ee

Given the fall-offs described in the appendices, for the vector field (\ref{newsymxi3}) one finds
\be
\rho_2=0,
\ee
 and we are only left with $\rho_1$. The computation of  $\rho_1$ is simplified due to the radiation gauge, the only terms contributing being:
\be
\rho^\soft = \rho_1= r^2 ( \frac{1}{2} \Gamma^t_{rr} \delta_\xi h^{rr} + \Gamma^t_{rA} \delta_\xi h^{rA} + \frac{1}{2}\Gamma^t_{AB}   \delta_{\xi} h^{AB} ).
\ee
From the expansions given in the appendices, one finds the $r^2$ and $r$ terms are:
\ba
\overset{(2)}{\rho} & = & \frac{1}{2} \partial_u C_{AB} D^A \xioneB \label{rhos1} \\
\overset{(1)}{\rho} & = & \frac{1}{2}( \partial_u \overset{(-1)}{h_{rA}}(\xioneA - \partial_u \xioA) 
+ \partial_u C_{AB} D^A \xioB + \partial_u \overset{(0)}{h_{AB}} D^A \xioneB)  \label{rhos2}
\ea
With the fall-offs (\ref{strongfallC}) these yield vanishing contributions to the $O(t^2)$ and $O(t)$ charges.\footnote{It is interesting to note that with the weaker fall-offs $C_{AB}(u,\xh) = O(|u|^{-1 - \epsilon})$ 
one obtains a nonzero $O(t)$ charge that corresponds to the `soft' part of the $O(t)$ hard charge found in Eq. (\ref{rho1hardX}). This is in compatibility with the prescription of Eq. (\ref{prescription}).}

Using Eq. (\ref{freehC}) the finite contributions of (\ref{rhos1}), (\ref{rhos2}) are found to be (discarding total derivatives in $u$ and in the sphere):
\be \label{divcont}
u^2  \overset{(2)}{\rho} -u \overset{(1)}{\rho} =  \C_{AB}(D^A \xioneB -\frac{1}{2} D^A  \partial_u \xioB - \frac{1}{4} \Delta D^A \xioneB), 
\ee
where we used that $\partial_u \xioneA=0$ and that  $\xioA$ is linear in $u$.

It remains to compute the $O(r^0)$ part of $\rho$.  This is found to be:
\ba
\overset{(0)}{\rho} &=&  \frac{1}{2}( - \partial_u \overset{(-1)}{h_{rA}} \partial_u \overset{(-1)}{\xi^A} + \partial_u \overset{(-2)}{h_{rA}}(\xioneA-\partial_u \xioA) + \partial_u C_{AB} D^A  \overset{(-1)}{\xi^B} + \partial_u  \overset{(0)}{h_{AB}} D^A \xioB +\partial_u  \overset{(-1)}{h_{AB}} D^A \xioneB ) \nonumber \\
&= & \C_{AB}(- D^A \xioneB +\frac{1}{2} D^A  \partial_u \xioB + \frac{1}{4} \Delta D^A \xioneB) + \frac{1}{2}\partial_u  \overset{(-1)}{h_{AB}} D^A \xioneB
\ea
where in the last line we  discarded total derivative terms. Here we used Eq. (\ref{freehC}) for all metric components except for $\overset{(-1)}{h_{AB}}$. The total finite charge is then:
\ba
\rho ^\soft &= &u^2  \overset{(2)}{\rho} -u \overset{(1)}{\rho}+ \overset{(0)}{\rho} \\ &= &\frac{1}{2}\partial_u  \overset{(-1)}{h_{AB}} D^A \xioneB \\
& =& \C_{AB} s^{AB} ,
\ea
where
\be \label{softterm}
s^{AB} = \frac{1}{16} \Delta^2 D^A \xioneB - \frac{3}{8} \Delta D^A \xioneB+ \frac{1}{2}D^A \xioneB.
\ee
In the last equality we used (\ref{freehC}), $D_A \xioneA=0$ and performed a few integrations by parts. 
When (\ref{softterm}) is expressed in $(z,\zb)$ coordinates, one finds (see appendix \ref{appsoft}):
\be \label{szz}
s_{zz}= - \frac{1}{4} D^4_z D^z \overset{(1)}{\xi^z}.
\ee
The soft charge can then be written as:
\be
\rho^\soft= \C^{zz} s_{zz} + c.c.
\ee
With the identification (\ref{Yitoxione}), this is  precisely  (-1/4 times) the soft charge proposed in Eq. (\ref{QsT1}).

\subsection{Summary}

As the previous discussion was rather dense with some tedious computations, here we summarize the main findings. We have shown that if we consider the  new class of large diffeomorphisms (\ref{newsymxi2}) which are parametrized by sphere vector fields $X^A \neq 0$ , $D_{A}X^A = 0$, and compute the associated (finite) charges via covariant phase space techniques, the corresponding Ward identities are implied by the sub-subleading soft theorem. 
We have thus reproduced ``one side of the equivalence"
between such symmetries and the soft theorem 
by showing that $\textrm{sub-subleading theorem}\ \implies\ \textrm{new symmetries Ward identities}$. The reason 
we do not yet have the converse ($\textrm{Ward identities}\ \implies\ \textrm{Sub-subleading theorem}$) is the following:
There are two sub-subleading theorems (for each angular direction at null infinity) associated to  positive  and negative helicity gravitons. However as the number of independent generators associated to new symmetries is only one  (due to $X^A$ being divergence-free), naively we have half the required number of charges/symmetries needed to 
reproduce the entire content of sub-subleading theorem. 

This tension has its antecedents in the equivalence between Weinberg soft theorem and Ward identities associated to supertranslation charges. Even in that case, one has two Weinberg soft theorems (for two polarization of soft gravitons) but only one charge associated to supertranslation vector fields which are parametrized by a single function. This tension was resolved by Strominger by using a remarkable condition \cite{strom1} which equated the amplitude for emitting a  positive helicity soft graviton with amplitude for emitting a negative helicity soft graviton, thereby reducing the  number of soft theorems to one. In \cite{strom1}
This condition arose from the fact that the perturbative gravity scattering processes can be thought of as weakly gravitating processes which preserve certain asymptotic conditions of the spacetime metric (originally derived by Christodoulou and Klainerman). However this condition only pertains to leading soft insertions and do not equate positive helicity insertions with negative helicity insertion, when the gravitons are sub-leading or sub-subleading. 
For the sub-leading theorems this is precisely what is desired as the associated Ward identities are generated by ``sphere" vector fields which have two independent components.  
Thus the questions remains, how does one derive two independent charges associated to the large diffeomorphisms considered in this paper which are parametrized by one function (divergence free vector field on the sphere). We do not answer this question in this paper but give a hint as to where the answer may lie. This hint itself presents a new perspective on the asymptotic charges by thinking of them in terms of electric and magnetic part of the Weyl tensor. 

In a nut-shell, in the following section we show how as far as supertranslation charges are concerned, for each supertranslation generator one has two independent charges! One is analogous to the electric charge in QED and the other one analogous to the magnetic charge. It is the gravitational electric charge, which is the supertranslation charge used in \cite{strom1}, whereas the vanishing of the  magnetic charge precisely gives the Christodoulou-Klainerman condition that we alluded to above.

\section{Electric and Magnetic charges for BMS} \label{sec7}
The structure of `soft photon' charges in QED \cite{maxwell} suggests there should be  `magnetic' dual charges to the canonical charges computed above. 
To support this idea, in this section we present a new way of interpreting generalized BMS charges as `electric' quantities with associated magnetic duals. We here departure from the main body of the paper in that the analysis is performed in the context of vacuum (non-linear) gravity in Bondi gauge.  Even though we expect the results should be derivable in de Donder gauge,  we do not attempt to do so in this paper.

The analysis presented in this section, together with the structure of `subleading' soft photon charges in QED suggests that the charges $\Q_Y$ found in section \ref{sec5} should be interpretable as  `electric' and `magnetic' charges associated to  $\xi^a \sim r X^A \partial_A$. We hope to be able to confirm this expectation in future investigations. 

\subsection{Electric charges}
In electrodynamics, the covariant phase space charges that generate gauge transformations can be written as \cite{leewald}:
\be \label{QSiglam}
Q_\Sigma[\lambda] = \int_{\Sigma}  \partial_a(\lambda  E^a),
\ee
where $\Sigma$ is a space-like Cauchy surface with normal $n^a$ and  $E^a= \sqrt{g} \, F^{a b} n_b$
the corresponding electric field. In \cite{maxwell} we used (\ref{QSiglam}) to obtain charges at null-infinity by taking the limit where $\Sigma$ approaches null infinity $\I$,
\be \label{Qscrilam}
Q_\I[\lambda]  = \lim_{\Sigma \to \I} Q_\Sigma[\lambda],
\ee
and in this manner recovered the charges  associated to the soft photon theorems.  Here we would like to find  an analogue  of (\ref{Qscrilam}) in gravity. 

The standard definition of gravitational  electric field (associated to the hypersurface $\Sigma$) is defined in terms of the Weyl tensor as:
\be \label{E1st}
E^{a}_{\ b} = - \sqrt{g}  \, C^{a c}_{\phantom{ac} b d} n_c n^d.
\ee
Based on how Poincare charges are expressed at spatial infinity \cite{hansen,virmani}, a first naive guess that generalizes (\ref{Qscrilam}) to gravity is then:
\be
Q_\I[\xi] \overset{?}{=}  \lim_{\Sigma \to \I} \int_{\Sigma}  \partial_a(  E^{a}_{\ b}  \xi^b). \label{QE1st}
\ee 
As explained below, this first guess needs two modifications in order to reproduce the required charges.

The first modification is well known: In order to get a non-trivial limit at null infinity one needs to rescale the Weyl tensor by an appropriate conformal factor \cite{aanull}. For our purposes, this will be achieved by including in  (\ref{E1st}) a factor of $r$. The second modification has to do with the null signature of the limiting surface $\I$: Since we are looking at vector fields $\xi^a$ that in the limit are tangent to $\I$, we want the index $b$ in (\ref{E1st}) to project along a direction that is transversal to $\I$. It is then natural  to consider projections along outgoing null directions. Thus, we will consider the contraction: $C^{a c}_{\phantom{ac} b d} n_c l^d$ where $l^a$ is an outgoing null vector. In Bondi gauge $l^a =\partial_r$ and so we propose a definition of electric field that in Bondi coordinates reads:
\be
\E^{a}_{\ b}\ := -  r \, \sqrt{g} \, C^{at}_{\phantom{at} br},
\ee
where we are considering $\Sigma$ to be a $t=u+r=$ constant hypersurface. 
One can verify (see appendix \ref{appE}) that  $\E^a_{\ b} = O(1)$ as $r \to \infty$. Thus, the `corrected' proposal takes the form:
\ba
Q_{\I}[\xi] & :=& \lim_{t \to \infty} \int_{\Sigma_t} \partial_a (\E^a_{\ b}\xi^b)  \label{QE1} \\
& =& \int_\I   \partial_u \big(  \overset{(0)}{\E^u_{\ u}}  \overset{(0)}{\xi^u} + \overset{(0)}{\E^u_{\ A}} \overset{(0)}{\xi^A} \big)  \label{QE2}
\ea
where in the last line we discarded a total sphere divergence. We now show that indeed (\ref{QE2}) reproduces the  generalized BMS group charges.

In Bondi gauge, the  electric field component at null infinity are found to be (see appendix \ref{appE}):
\ba
\overset{(0)}{\E^u_{\ u}} &  =& \sqrt{q}(-2 M + \frac{1}{4} \partial_u \bo \label{E0uu} ) \\
\overset{(0)}{\E^u_{\ A}} & = & \sqrt{q}(-N_A + 3 \, \partial_A \bo) \label{E0uA}
\ea
where $N_A$ and $M$ are the momentum and mass aspects and  $\bo = -\frac{1}{32} C^{AB} C_{AB}$. For a supertranslation vector field $\xi_f^a = f \partial_u$ expression (\ref{QE2}) becomes:
\be
Q_\I[\xi_f] = -2 \int_\I  \sqrt{q} f \partial_u M , \label{QEf}
\ee
where the piece associated to the second term in (\ref{E0uu}) integrates to zero with the standard fall-offs $C_{AB}(u)= C^{\pm}_{AB} +O(|u|^{-\epsilon})$.  The expression coincides with the radiative space supertranslation charge \cite{as}. One can also check (see appendix \ref{weylvfcharges}) that for generalized BMS vector field  $\xi_V^a  =  V^A \partial_A + u \alpha \partial_u$ the charge coincides with the one obtained in \cite{cl2} by covariant phase space methods.

\subsection{Magnetic charges}
In analogy to the QED case, we propose to define the magnetic `dual' charges of (\ref{QE1}) as
\be
Q^*_{\I}[\xi]  = \lim_{t \to \infty} \int_{\Sigma_t} \partial_a (\B^a_{\ b}\xi^b)  \label{QB1}
\ee
where
\be
\B^{a}_{\ b}\ :=  -  r \, \sqrt{g} \, *C^{at}_{\phantom{at} br} 
\ee
and $*C^{at}_{\phantom{at} br} \equiv \frac{1}{2} \epsilon^{a t c d} C_{c d b r}$. The leading components of the magnetic field are (see appendix \ref{magweyl}): 
\ba
\overset{(0)}{\B^u_{\ u}} &  =& \frac{\sqrt{q}}{2} \epsilon^{AB} (D_B D^M C_{AM} + \frac{1}{2} \dot{C}_{AM} C^M_B)   \label{B0uu} \\
\overset{(0)}{\B^u_{\ A}} & = &-\epsilon_A^{\phantom{A}B} \overset{(0)}{\E^u_{\ B}}  \label{B0uA}.
\ea
Thus, for a supertranslation vector field the charge (\ref{QB1}) becomes
\be
Q^{*}_\I[\xi_f] = \frac{1}{2} \int_\I  \sqrt{q} f \epsilon^{AB} D_B D^M \dot{C}_{AM} 
\ee
(the contribution coming from the second term in (\ref{B0uu}) integrates to zero). The vanishing of the magnetic charge corresponds to the  Christodoulou-Klainerman (CK) condition  \cite{strom1}. Whence for each supertranslation generator, there are two charges, one arising from electric part of Weyl tensor and the other from the magnetic part of Weyl tensor. The vanishing of magnetic charge  implies that positive and negative soft insertions are equal to each other, and then the Ward identities associated to electric charge implies Weinberg's Soft theorem.

In appendix \ref{weylvfcharges} we comment on the magnetic charges associated to sphere vector fields. 
\section{Summary and open issues}
In gauge theories as well as gravity, we have a hierarchy of soft theorems, many of whom have been interpreted as Ward identities asssociated with spontaneously broken symmetries.  Up untill this point,  the sub-subleading soft graviton theorem was lacking such interpretation.  In this work, we have proposed just such an interpretation to the sub-subleading soft graviton theorem.

We started by `reading off' candidate charges from the soft theorem expression, following \cite{stromst,stromlow}. From this analysis, given in section \ref{sec5}, 
one concludes that the sub-subleading soft theorem is equivalent to statement that the S matrix commutes with certain charges $\Q_Y$,
\be
[\Q_Y ,S ] =0 \iff \text{sub-subleading CS soft theorem},
\ee
where the charges are parametrized by symmetric, trace-free tensors on the 2-sphere $Y^{AB}$. They are the gravitational analogue of the charges found for QED in \cite{stromlow}, which were parametrized by vector fields on the sphere.  Having found the charges from the soft theorem, our next goal was to derive them from first principles.
Based on an analogue derivation 
in QED \cite{maxwell}, we set out to explore asymptotic charges associated to vector fields that are more general than the so far considered generalized BMS.  By demanding IR divergences to be controlled in the way spelled out in section \ref{sec6}, we found a new set of vector fields with asymptotic form
\be \label{eq1conc}
\xi^a \sim r X^A \partial_A, \quad D_A X^A =0,
\ee
whose associated (finite) charges  $Q_X$  correspond to the charges $\Q_Y$. Specifically, we showed
\be
Q_{X^A} = -\frac{1}{4} \Q_{(D^A X^B)^{\stf}}.
\ee
However, due to the divergence free condition of $X^A$ (\ref{eq1conc}), the charges $Q_X$ do not exhaust all possible $\Q_Y$ charges. 
To see what is missing, recall every symmetric, trace-free tensor can be decomposed as
\be
Y^{AB}= (D^A X^B + \epsilon^{B}_{\;C} D^A X'^C )^{\stf}, \quad D_A X^A = D_A X'^{A}=0.
\ee
From this decomposition it becomes clear that we have only  recovered `half' of the charges $\Q_Y$. We expect that the `remaining half' is associated to a `magnetic-dual' charge, in analogy to the QED case \cite{maxwell}. To support this idea, we showed in section \ref{sec7} how there exists a natural casting of supertranslation charges in terms of the electric part of the Weyl tensor. We then saw how, upon dualizing the Weyl tensor, the resulting expression yields the `magnetic supermomentum' charge \cite{nut4} that appears implicitly in the analysis of asymptotic symmetries and Weinberg's soft graviton theorem \cite{stromst} (see \cite{prl} for a lengthier discussion).  However, extending this analysis to the current `sub-subleading' is left for future investigations. 

There are many open issues that arise out of this current work in addition to the one mentioned above.  We outline some of them below.\\
{\bf (a)}  Perhaps the most pertinent  question is the precise meaning of these large gauge transformations. Whereas  generalized BMS can be understood as a group that maps an asymptotically flat spacetime to another asymptotically flat spacetime, here we do not even have a group to begin with! (the vector fields (\ref{eq1conc}) do not close under vector field commutator). Is there any sense in which they can be thought of as (classical) symmetries of Einstein's equations?\\
{\bf (b)} Is there any physical/geometrical interpretation of the charges $\Q_Y$? \\
{\bf (c)} From the scattering amplitude side, it seems that the soft graviton factorization stops at sub-subleading order \cite{loopcorr}. Can this be understood from the covariant phase space perspective (as for instance argued in  \cite{maxwell} for the absence of sub-subleading factorization in QED)?\\
{\bf (d)} As the fate of both the sub and sub-subleading theorems is not settled once loop corrections are taken into account, at most the diffeomorphisms we have considered in this paper are symmetries of tree-level (semi-classical) gravity. It is unclear what their fate will be in quantum gravity.\\
{\bf (e)} Our analysis in this paper strictly pertains to perturbative gravity. Are the sub-subleading soft theorems also associated to symmetries of fully non-linear semi-classical gravity? \\

\appendix
\section{Minkowski metric and differential operators in retarded coordinates}
Minkowski metric in retarded coordinates $u=t-r$, $r$ and $x^A$, $A=1,2$ is given by
\be
ds^2 = -du^2 -2 du dr +r ^2 q_{AB}dx^A dx^B
\ee
with $q_{AB}$ the unit sphere metric. The nonzero Christoffel symbols are:
\be
\Gamma^A_{r B}= r^{-1} \delta^{A}_B, \quad \Gamma^{r}_{AB}= -r q_{AB}, \quad \Gamma^u_{AB}= r q_{AB}.
\ee
For sphere derivatives we use the covariant derivative $D_A$ compatible with $q_{AB}$ and so the Christoffel symbols $\Gamma^{A}_{BC}$ do not appear explicitly.

The wave operator acting on a vector field takes the form:
\ba
r \,  \square \, \xi^r &  =&  \partial_r^2( r \xi^r) - 2 \partial_u \partial_r (r \xi^r) + r^{-1} (\Delta - 2)\xi^r - 2 D_A \xi^A  \nonumber  \\
r \,  \square \, \xi^u &  =&  \partial_r^2( r \xi^u) - 2 \partial_u \partial_r (r \xi^u) + r^{-1} \Delta \xi^u
+ 2 r^{-1} \xi^r + 2 D_A \xi^A \label{boxxiru} \\
r^2 \,  \square \, \xi^A &  =& \partial_r^2(r^2 \xi^A) - 2 \partial_u \partial_r (r^2 \xi^A)+ (\Delta-1)\xi^A +2 r^{-1} D^A \xi^r \nonumber 
\ea
where $\Delta = D_A D^A$ is the Laplacian on the sphere. 

\section{Vector field and related expansions}
\subsection{Vector field} \label{appboxxi}
The wave equation (\ref{boxxiru}) applied to the  ansatz (\ref{newsymxi}) yields the equations to be satisfied by the coefficients of the $r \to \infty$ expansion. The vanishing of the leading term yields:
\be
\partial_u \xitwor =0, \quad \quad  \partial_u \xitwou=0, \quad \quad \partial_u \xioneA =0.
\ee
For the next terms one finds:
\be
\begin{array}{ll}
r \,  \square \, \xi^r =& r( -4 \partial_u \xioner +(\Delta +4) \xitwor -2 D_A \xioneA) +O(1) \\
r \,  \square \, \xi^u =& r(-4 \partial_u \xioneu+ 2 \xitwor +(\Delta + 6) \xitwou + 2 D_A \xioneA) \\
&  \quad\quad\quad +  (-2 \partial_u \xiou + 2 \xioner +(\Delta+2) \xioneu+ 2 D_A \xioA)+O(r^{-\epsilon})  \\
r^2 \,  \square \, \xi^A  =&  r (-4 \partial_u \xioA +(\Delta +5) \xioneA + 2 D^A \xitwor )+O(1)
\end{array}
\ee
Conditions (\ref{xi2ueq0}), (\ref{logcond}) together with the $O(r)$ condition of $r \square \xi^u=0$ impily
\be
\xitwor=0, \quad \quad D_A \xioneA=0.
\ee
This in turn implies $\partial_u \xioner=0$. Let us set this `integration constant' as:
\be
\xioner = -\alpha(\xh)
\ee
The equation for $\xioA$ gives
\be
\xioA = \frac{u}{4}(\Delta+5) \xioneA + V^A(\xh)
\ee
with $V^A$ an `integration constant'. 
Finally, the $O(1)$ condition for $r \square \xi^u=0$ gives
\be
\xiou= u(- \alpha + D_A V^A) + f(\xh)
\ee
with $f$ an `integration constant'. $f$ is associated to supertranslations and $\alpha$ and $V^A$ to `subleading' vector fields.
Hence for the purpose of the sub-subleading charges, we set all these integration `constants' to zero. The resulting vector field has the form given in (\ref{newsymxi3}).
 
\subsection{$\delta_\xi h^{ab}$, etc}
In retarded coordinates, $\delta_\xi h^{ab}= \nabla^a \xi^b + \nabla^b \xi^a$ is given by:
\ba
\delta_\xi h^{rr} & = & 2(\partial_r -\partial_u) \xi^r \\
\delta_\xi h^{ru} & = & \partial_r(\xi^u-\xi^r) - \partial_u \xi^u \\
\delta_\xi h^{uu} &=& -2 \partial_r \xi^u \\
\delta_\xi h^{rA} &=& (\partial_r -\partial_u)\xi^A+r^{-2}D^A \xi^r \\
\delta_\xi h^{uA} &=& -\partial_r \xi^A + r^{-2} D^A \xi^u \\
\delta_\xi h^{AB}&=& r^{-2} (D^A \xi^B + D^B \xi^A) + 2 r^{-3}q^{AB} \xi^r.
\ea
For the vector field (\ref{newsymxi}) this gives the following leading $r \to \infty$ terms: 
\be \nonumber
\begin{array}{l}
\overset{(1)}{\delta_\xi h^{rr}}  = 4 \xitwor-2\partial_u \xioner \\
 \overset{(1)}{\delta_\xi h^{ru}}  =  2\xitwou -2 \xitwor-\partial_u \xioneu, \quad 
  \overset{(0)}{\delta_\xi h^{ru}}= \xioneu-\xioner - \partial_u \xiou \\
\overset{(1)}{\delta_\xi h^{uu}} = -2 \xitwou, \quad \quad \overset{(0)}{\delta_\xi h^{uu}} = -2 \xioneu \\
\overset{(0)}{\delta_\xi h^{r A}}= \xioneA-\partial_u \xioA + D^A \xitwor, \quad
\overset{(-1)}{\delta_\xi h^{r A}}= -\partial_u \overset{(-1)}{\xi^A} + D^A \xioner \\
\overset{(0)}{\delta_\xi h^{u A}} = - \xioneA +D^A \xitwou, \quad \quad \overset{(-1)}{\delta_\xi h^{u A}} =D^A \xioneu \\
\overset{(-1)}{\delta_\xi h^{AB}} = D^A \overset{(1)}{\xi^B} + D^B \overset{(1)}{\xi^A} + 2 q^{AB} \xitwor , \quad \quad \overset{(-2)}{\delta_\xi h^{AB}} = D^A \overset{(0)}{\xi^B} + D^B \overset{(0)}{\xi^A} + 2 q^{AB} \xioner \\
\overset{(-3)}{\delta_\xi h^{AB}} = D^A \overset{(-1)}{\xi^B} + D^B \overset{(-1)}{\xi^A}  + 2 q^{AB} \xior
\end{array}
\ee 
The divergence of the vector field has the expansion:
\be
\nabla_a \xi^a = r \overset{(1)}{\nabla_a \xi^a} + \overset{(0)}{\nabla_a \xi^a} + O(r^{-\epsilon})
\ee
with
\ba
\overset{(1)}{\nabla_a \xi^a}& = & 4 \xitwor + \partial_u \xioneu +D_A \xioneA \\
\overset{(0)}{\nabla_a \xi^a}  &= & 3 \xioner+ \partial_u \xiou+ D_A \xioA.
\ea
For $\rho_3$ we need  the following components of
 $\delta_\xi \hb^{ab} = \delta_\xi h^{ab} - \nabla_c \xi^c \eta^{ab}$:
 \ba
\delta_\xi \hb^{tr} & =& \delta_\xi h^{rr} + \delta_\xi h^{ur} \\
\delta_\xi \hb^{tu}& = & \delta_\xi h^{ru}+ \delta_\xi h^{uu} + \nabla_c \xi^c\\
\delta_\xi \hb^{tA}& = & \delta_\xi h^{rA} + \delta_\xi h^{uA}
\ea

\subsection{$\delta_\xi \Gamma^t_{ab}$}
From the expression of the Christoffel symbols one can verify the identity  $\delta_\xi \Gamma^t_{ab}= \nabla_a \nabla_b \xi^t$. In components this gives:
\ba
\delta_\xi \Gamma^t_{rr} & = & \partial^2_r \xi^t = 2 (\xitwor +\xitwou) + O(r^{-2-\epsilon})  \\
\delta_\xi \Gamma^t_{ru} &= & \partial_r \partial_u \xi^t = \partial_u (\xioner+\xioneu)+ O(r^{-1-\epsilon})\\
\delta_\xi \Gamma^t_{uu} &= & \partial^2_u \xi^t =O(1)\\
\delta_\xi \Gamma^t_{rA} &=& r \partial_r (r^{-1} D_A \xi^t) = r D_A(\xitwor+\xitwou)+O(r^{-1}) \\
\delta_\xi \Gamma^t_{uA} &=& D_A \partial_u \xi^t = O(r) 
\ea
\begin{multline}
\delta_\xi \Gamma^t_{AB} = D_A D_B \xi^t + r q_{AB}(\partial_r -\partial_u) \xi^t  =
 r^2( D_A D_B \xitwot +q_{AB}(2 \xitwot - \partial_u \xionet) \\
 +r(D_A D_B \xionet +q_{AB}(\xionet-\partial_u \xiot)) +O(1)
\end{multline}
where we used that $\xi^t=\xi^r+\xi^u$ and considered a general vector field of the type (\ref{newsymxi}).

\section{Free metric perturbation} \label{appfreeh}
For the free metric perturbation, we seek for an asymptotic solution to the linearized vacuum Einstein equations with given free data $C_{AB}$. After imposing de Donder gauge one can still use residual gauge transformation to further restrict the metric components. Here we will use `radiation gauge'  (see e.g. section 4.4b of \cite{wald}) which in retarded coordinates reads
\be \label{radgauge}
h_{a u}=0, \quad \eta^{ab} h_{ab}=0.
\ee
Thus, we seek for asymptotic solutions to 
\be \label{eqnsfreeh}
\square h_{ab} =0, \quad \nabla^b h_{ab}=0
\ee
with metric perturbations of the form (\ref{radgauge}). Assuming standard  $1/r$ expansion and imposing compatibility with (\ref{eqnsfreeh}) one is lead to the following  fall-offs:
\be
h_{rr} = O(r^{-3}), \quad h_{rA} = O(r^{-1}) , \quad h_{AB} = r \, C_{AB} +O(1).
\ee
Assuming a $1/r^n$ expansion,  equations (\ref{eqnsfreeh}) can then be solved iteratively. The leading terms relevant for this paper are found to be:
\be \label{freehC}
\begin{array}{llr}
\partial_u \overset{(-1)}{h_{Ar}} = D^B C_{AB}  & & \partial_u \overset{(0)}{h_{AB}} =(-\frac{1}{2}\Delta +1) C_{AB} \\
\partial_u \overset{(-2)}{h_{Ar}} =\overset{(-1)}{h_{Ar}}+ D^B \overset{(0)}{h_{AB}} & &
\partial_u \overset{(-1)}{h_{AB}} = -\frac{1}{4}\Delta \overset{(0)}{h_{AB}} -D_{(A}  \overset{(-1)}{h_{B)r}}\\
 \partial_u  \overset{(-3)}{h_{rr}} = D^B  \overset{(-1)}{h_{Br}} &&
\end{array}
\ee
From these expressions one can obtain the components of the contravariant metric perturbation and linearized Christoffel symbols.  The nonzero components are:
\ba
h^{rr}=h^{uu}=-h^{ru}=h_{rr} = O(r^{-3}), \nn \\
 h^{rA}=-h^{uA} =r^{-2}q^{AB} h_{rB} = O(r^{-3}), \\
 h^{AB}= r^{-4}q^{AM}q^{BN} h_{MN} = O(r^{-3}) \nn
\ea
and
\ba
\Gamma^t_{rr} & =& \frac{1}{2}\partial_u h_{rr} = O(r^{-3}) \nn \\
\Gamma^t_{rA} & =& \frac{1}{2}\partial_u h_{rA} = O(r^{-1}) \\
\Gamma^t_{AB} & =& \frac{1}{2}\partial_u h_{AB} = O(r) .\nn
\ea 
\section{Stress tensor expansion} \label{stressapp}
The free scalar field $\varphi$ has an expansion
\be \label{expphi}
\varphi = r^{-1} \phi  + r^{-2} \phitwo +O(r^{-3})
\ee
where $\phi$ is the free data. In particular from $\square \phi =0$ the subleading term is determined according to
\be
\partial_u  \phitwo  = -\frac{1}{2} \Delta \phi. \label{phitwo}
\ee
The stress tensor
\be
\T_{ab} = \partial_a \varphi \partial_b \varphi - \frac{1}{2}\eta_{ab} |\nabla \varphi|^2.
\ee
is then found to have the following fall-offs
\be
\begin{array}{c}
\T_{rr}=O(r^{-4}), \quad \T_{u r}  =  O(r^{-4}) , \quad \T_{uu}   = O(r^{-2})\\
\T_{rA}  = O(r^{-3}), \quad \T_{uA}  = O(r^{-2}) \\
\T_{AB}=O(r^{-1})
\end{array}
\ee
The leading components can then be easily computed. For instance:
\ba
\Ttwo_{uu} &= & \phid^2 \\
\Ttwo_{uA} & = & \phid \,  \partial_A \phi \\
\Tthree_{u A} &  = &  \partial_A \phi \, \partial_u \phitwo + \partial_A \phitwo \partial_u \phi \label{Tm3uA}
\ea
For the most divergent vector fields used in the paper, where $\xi^A=O(r)$, $\xi^r=O(r)$ and $\xi^u=O(r^0)$, the stress tensor contribution to the hard charge, (\ref{rhoT}), is given by:
\be \label{rhoTasym}
\rho_T \equiv - r^2 \T^t_{\, a} \xi^a =  r \big( \xioneA \Ttwo_{uA}\big)  + \big( \xioneA \Tthree_{u A} + \xioA\Ttwo_{u A}  + \xiou \Ttwo_{uu} \big) + O(r^{-\epsilon})
\ee

\subsection{Eq. (\ref{T3uA})} \label{appT3uA}
Using (\ref{Tm3uA}) and (\ref{phitwo}), and discarding total $u$ and sphere derivatives one finds:
\be
\overset{1}{\xi^A} \Tthree_{u A} = - \overset{1}{\xi^A} D_A \phi \, \Delta \phi - \frac{1}{2} D_B \overset{1}{\xi^B} \Delta \phi \, \phi \label{T3uA2}
\ee
Now using the identities:
\ba
V^A D_A \Delta \phi =  -D^{(A} V^{B)} (D_A \phi D_B \phi)^{\tf} + D_B(V^A D_A\phi D^B \phi - \frac{1}{2}|D \phi |^2 V^B ) \\
f \Delta \phi \, \phi   =  \frac{1}{2} \Delta f \phi^2 - f |D \phi|^2 +D_A(f \phi D^A \phi - \frac{1}{2} D^Af \phi^2) 
\ea
one can express (\ref{T3uA2}) in terms of factors that only contain single derivatives of $\phi$ plus total  derivatives. For the case of divergence free vector field the only term that survives is the one given in Eq. (\ref{T3uA}).

\section{Sourced metric perturbation expansion}\label{appsourcedh}
In this section we describe the asymptotic solution for the sourced (trace-reversed) metric perturbation $\hb_{ab}$,
\be \label{eqnssourcedh}
\square \hb_{ab} = - 2 \T_{ab}, \quad \nabla^b \hb_{ab}=0
\ee
with $\T_{ab}$ as given in the previous section.   By looking at these equations for $r \to \infty$ one is led to consider the following leading nonzero orders:
\be \label{fallsourcedh}
\begin{array}{c}
\hb_{rr}= O(r^{-3} \ln r), \quad \hb_{ru}= O(r^{-2} \ln r), \quad \hb_{uu}= O(r^{-1} \ln r), \\
\hb_{rA} = O(r^{-2} \ln r), \quad \hb_{uA} = O(r^{-1} \ln r), \quad \hb_{AB} = O(\ln r)
\end{array}
\ee
Assuming an expansion in $1/r^n$ and $\log r/r^n$, one can solve the equations  (\ref{eqnssourcedh}) at each order. It is convenient to  express the solution in terms of:
\be
\mu:= \sint^u (\partial_u \phi)^2, \quad  \phi^2 , \quad  \Tthree_{uu}, \quad T_{uA}:=\Ttwo_{uA} , \quad (T_{AB})^{\tf}:=(\Ttwo_{AB})^{\tf}.
\ee
The leading terms are found to be:
\be \nn
\begin{array}{c}
\overset{-3 \ln}{\hb_{rr}}= 2 \sint^u \sint^{u'} \mu , \quad \overset{-3}{\hb_{rr}}= \sint^u \sint^{u'}\big( 3 \mu + 2 \overset{-1}{\hb_{uu}} \big) -\sint^u \phi^2 \\
 \overset{-2 \ln}{\hb_{ru}}= - \sint^u \mu, \quad  \overset{-2 }{\hb_{ru}}= - \sint^u( \mu+\overset{-1}{\hb_{uu}}), \quad
 \overset{-3 \ln}{\hb_{ru}} = \sint^u \Delta \mu \\
\overset{-1 \ln}{\hb_{uu}}= \mu, \quad  \overset{-2 \ln}{\hb_{uu}}= -\frac{1}{2} \sint^u \Delta \mu \quad \overset{-2}{\hb_{uu}}= \frac{1}{2} \sint^u \big( \mu- \Delta \mu - \Delta \overset{-1}{\hb_{uu}} -2 \overset{-3}{\T_{uu}}\big)
\\
 \overset{-2 \ln}{\hb_{rA}}= -2 \sint^u \sint^{u'} D_A  \mu, \quad  
  \overset{-2 }{\hb_{rA}}= \sint^u \sint^{u'}\big(  D_A( -3  \mu -2 \overset{-1}{\hb_{uu}}+ \phid \phi) + T_{uA} \big) \\
  
 \overset{-1 \ln}{\hb_{uA}}= \sint^u  D_A  \mu, \quad  \overset{-1 }{\hb_{uA}} = \sint^u \big(  D_A(\mu+ \overset{-1}{\hb_{uu}}) -T_{uA} \big) ,\quad 
  \overset{-2 \ln}{\hb_{uA}} = - \frac{1}{2} \sint^u \sint^{u'}  D_A \Delta \mu  \\
  \overset{-2 }{\hb_{uA}} = \sint^u \sint^{u'}\big(\frac{1}{4} (\Delta+1)T_{uA} - \frac{1}{2} \overset{-3}{\T_{uA}} \big) +D_A(\ldots)  \\
 \overset{\ln}{\hb_{AB}}= -\mu \, q_{AB} , \quad  \overset{0}{\hb_{AB}}= q_{AB}\big( \frac{1}{2}\phi^2 - \sint^u( \mu+ \overset{-1}{\hb_{uu}}) \big)\\
  \overset{-1 \ln}{\hb_{AB}} = \sint^u \sint^{u'}\big( D_A D_B  \mu + q_{AB}( \frac{1}{2} \Delta \mu - \mu) \big) \\
  \overset{-1 }{\hb_{AB}} = -\sint^u D_{(A} T_{B)u}-\frac{1}{2}T_{AB} +q_{AB}(\ldots ) +D_A D_B(\ldots)
\end{array}
\ee
From here one can obtain all the relevant metric dependent quantities.  For the trace  $\hb = \eta^{ab}\hb_{ab}$ one finds:
\be
\overset{-2 \ln}{\hb}= 0, \quad \overset{-2 }{\hb}= \phi^2, \quad   \overset{-3 \ln}{\hb}= 0
\ee
We note that the term $\overset{(-1)}{\hb_{uu}}$ is undetermined by the equations and corresponds to a `pure gauge' solution. As expected, this term does not feature in the charges.

\subsection{Christoffel symbols $\Gamma^t_{ab}$ sourced metric}
In terms of the traced-reversed metric perturbation $\hb_{ab}$, the linearized Christoffel symbols $\Gamma^t_{ab} =\Gamma^r_{ab}+\Gamma^u_{ab}$ read:
\be
\begin{array}{l}
\Gamma^t_{rr}=-\partial_r \hb_{ru}+\frac{1}{2} \partial_u \hb_{rr}-\frac{1}{2}\partial_r \hb,\quad
\Gamma^t_{ru}=-\frac{1}{2}\partial_r \hb_{uu}-\frac{1}{4}\partial_r \hb, \quad \Gamma^t_{uu}=-\frac{1}{2}\partial_u \hb_{uu}-\frac{1}{4}\partial_u \hb \\
\\
\Gamma^t_{rA}=-\frac{1}{2}D_A(\hb_{ru}+\frac{1}{2}\hb) -\frac{1}{2} r^2 \partial_r(r^{-2} \hb_{uA})+\frac{1}{2}\partial_u \hb_{rA} , \quad
\Gamma^t_{uA}= -\frac{1}{2} D_A(\hb_{uu}+\frac{1}{2} \hb) \\
\\
\Gamma^t_{AB}= - D_{(A} \hb_{B)u}+\frac{1}{2} \partial_u \hb_{AB}+q_{AB}\big( r(\hb_{uu}-\hb_{ru}) -\frac{1}{4} r^4 \partial_u \hb\big).
\end{array}
\ee
From the metric perturbation expansion given in the previous section we find the following leading nonvanishing terms:
\be
\begin{array}{l}
\overset{-3 \ln}{\Gamma^t_{rr}} = -\sint^u \mu, \quad \overset{-3 }{\Gamma^t_{rr}}=\sint^u (-\overset{(-1)}{\hb_{uu}}+\frac{1}{2}\mu) +\frac{1}{2}\phi^2  \\
\\
\overset{-2 \ln}{\Gamma^t_{ru}}= \frac{1}{2} \mu, \quad \overset{-2}{\Gamma^t_{ru}} =  \frac{1}{2} ( \overset{(-1)}{\hb_{uu}} - \mu ) \\
\\
\overset{-1 \ln}{\Gamma^t_{uu}}= - \frac{1}{2} \partial_u \mu \\
\\
\overset{-2 \ln}{\Gamma^t_{rA}}= \sint^u D_A \mu , \quad  \overset{-2 }{\Gamma^t_{rA}}= \sint^u(- T_{uA} + D_A \overset{(-1)}{\hb_{uu}} )\\
\\
\overset{-1 \ln}{\Gamma^t_{uA}}= -\frac{1}{2} D_A \mu, \quad \overset{-2 \ln}{\Gamma^t_{uA}}= \frac{1}{4} \sint^u D_A \Delta \mu , \quad \overset{-1}{\Gamma^t_{uA}}= -\frac{1}{2} D_A \overset{(-1)}{\hb_{uu}} \\
\\
\overset{\ln}{\Gamma^t_{AB}}= \frac{1}{2} q_{AB} \mu , \quad 
\overset{-1 \ln}{\Gamma^t_{AB}}= \sint^u\big( -\frac{1}{2}D_A D_B \mu +\frac{1}{2} q_{AB}(-\frac{1}{2} \Delta  \mu + \mu)\big) \\
\\
\overset{0}{\Gamma^t_{AB}}= \frac{1}{2} q_{AB} ( \overset{(-1)}{\hb_{uu}} - \mu )
\end{array}
\ee

\section{Eq (\ref{szz})} \label{appsoft}
By expressing derivatives of $X^z$ in terms of divergences and laplacians, one can show the identity:
\be
D^4_z D^z X^z = D^2_z( \frac{1}{2} \Delta D \cdot X + D\cdot X) -\frac{1}{4}\Delta^2 D_z X_z + \frac{3}{2} \Delta D_z X_z -2 D_z X_z
\ee
for any sphere vector field $X^A$. For divergence-free $X^A$, the first term vanishes and the expression corresponds with what is found in (\ref{softterm}).

\section{Electric and Magnetic parts of Weyl tensor at infinity in Bondi gauge} \label{appE}
\subsection{Electric part of Weyl}
We follow \cite{barnich} for the expression of the metric and Christoffel symbols in Bondi gauge. Rearranging indices and using 
\be \label{sqrtg}
\sqrt{g}=r^2 \sqrt{q} e^{2 \beta},  \quad g^{ur}= -e^{-2 \beta} 
\ee
we have
\be \label{Eua}
\E^{u}_a := - r \sqrt{g} \, C^{ut}_{\phantom{ut}ar} =  r^3 \sqrt{q} \,C_{arr}^{\phantom{aaa}r} .
\ee
Given the expressions of  Christoffel symbols in Bondi gauge \cite{barnich}, one finds:
\ba
 C_{urr}^{\phantom{aaa}r} & =&  \partial_r \Gamma^r_{ur}- \partial_u \Gamma^{r}_{rr} + \Gamma^B_{ur}\Gamma^r_{Br} \\
&=& r^{-3}( -2 M + 4 \partial_u \bo) + O(r^{-4}) \label{Currr}
\ea
where 
\be
\bo= - \frac{1}{32} C^{AB}C_{AB}  
\ee
and $M$ the Bondi mass aspect that satisfies \cite{barnich}:
\be \label{partialuM}
\partial_u M = -\frac{1}{8} \dot{C}^{A}_B\dot{C}^{B}_A +\frac{1}{4}D_A D_B \dot{C}^{AB}
\ee
Multiplying (\ref{Currr}) by $r^3 \sqrt{q}$ we obtain $\overset{(0)}{\E^u_{\ u}} $ as given in Eq. (\ref{E0uu}). For $a=A$ one finds:
\ba
C_{Arr}^{\phantom{aaa}r} & = &  r^{-1}\partial_r(r \Gamma^r_{Ar} )-\partial_A \Gamma^r_{rr}+\Gamma^B_{Ar} \Gamma^r_{r B} \\
& = & r^{-3}( -N_A + 3 \partial_A \bo) +O(r^{-4}) ,\label{CArrr}
\ea
which upon multiplying  by $r^3 \sqrt{q}$ gives Eq. (\ref{E0uA}). Here $N_A$ is the `angular momentum aspect' that satisfies \cite{barnich}:
\be \label{angmomasp}
-\partial_u N_A + 3 \partial_A \partial_u \bo= -\partial_A M + H_A + S_A
\ee
with
\ba \label{EAh}
H_A & =&  - \frac{1}{4} \partial_A (\dot{C}^{MN} C_{MN})
+\frac{1}{4} \dot{C}^{N}_M D_A C^M_N  
 + \frac{1}{4} D_B(C^B_M \dot{C}^M_A-\dot{C}^B_M C^M_A) \\
S_A & = &\frac{1}{4} D_B  (D_A D_M C^{BM} - D^B D_M C^{M}_{A}) 
\ea
(the notation for these piece stands for `Hard' and `Soft').  

\subsection{Magnetic part of Weyl}\label{magweyl}
Our starting point is the `dual' of Eq. (\ref{Eua}),
\be
\B^{u}_a =  r^3 \sqrt{q} \,*C_{arr}^{\phantom{aaa}r} \label{Bua}
\ee
Taking $a=u$  we find
\ba
*C_{urr}^{\phantom{aaa}r} & = & \frac{1}{2} \epsilon_{ur AB} C^{AB\phantom{r} r}_{\phantom{AB} r} \\
& = & r^{-3} \frac{1}{2} \epsilon^{AB} \overset{(-1)}{C_{ABr}^{\phantom{ABr}r}} +O(r^{-4}),
\ea
where we used Eq. (\ref{sqrtg}) and $\epsilon_{ur AB}= e^{2 \beta} r^{2} \epsilon_{AB}$ with $\epsilon_{AB}$ the area form of the unit sphere. Computing $C_{ABr}^{\phantom{ABr}r}$ and substituting in (\ref{Bua}) one obtains (\ref{B0uu}).

For $a=A$ we use $\epsilon_{Ar Bu}= e^{2 \beta} r^{2} \epsilon_{AB}$ and Eq. (\ref{sqrtg}) to  obtain
\ba
*C_{Arr}^{\phantom{aaa}r} & =& - r^2 \epsilon_{AB} g^{BM} C_{Mrr}^{\phantom{rrr}r} \\
& =& - r^{-3} \epsilon_{A}^{\phantom{A} B} \overset{(-3)}{C_{Brr}^{\phantom{rrr}r}} +O(r^{-4})
\ea
Comparing with (\ref{Eua}) we arrive at Eq. (\ref{B0uA}).

\subsection{Sphere vector field charges} \label{weylvfcharges}
For $\xi_V^a  =  V^A \partial_A + u \alpha \partial_u$ the electric charge takes the form:
\be \label{rhoV}
Q_\I[\xi_V] = \int_\I \rho_V \quad  \text{with} \quad  \rho_V  :=    V^A \partial_u \overset{(0)}{\E^u_A} + \alpha \, \partial_u (u \, \overset{(0)}{\E^u_u} ) .
\ee
As in \cite{cl2}, the charge is only  well-defined in the subspace of free data satisfying  the stronger fall-offs $C_{AB} = O(u^{-1-\epsilon})$. Substituting (\ref{E0uu}), (\ref{E0uA}) and (\ref{angmomasp}) in (\ref{rhoV}) we have (in the following we omit multiplicative $\sqrt{q}$ factors):
\be
\rho_V = V^A( -\partial_A M + H_A +  S_A) - 2 \alpha M - 2 \alpha u  \partial_u M + \frac{\alpha}{4} \partial_u(u \partial_u \bo)
\ee
The first and fourth terms add up to a total sphere divergence, and the last term is a total $u$ derivative that  does not contribute to the charge. Separating `hard' and `soft' contributions we have $\rho_V = \rhoh_V + \rhos_V$ with
\be \label{rhoVhs}
\rhoh_V = V^A H_A -2 \alpha u \partial_u M^{\text{hard}} , \quad  \rhoh_V = V^A S_A -2 \alpha u \partial_u M^{\text{soft}}.
\ee
where the `hard' and `soft' piece of $\partial_u M$ are the first and second term in (\ref{partialuM}) respectively. Combining all terms and discarding total derivatives one finds
\be
\rhoh_V = \frac{1}{4} \dot{C}^{AB}(\L_V C_{AB} + \alpha u \dot{C}_{AB})
\ee
\be
\rhos_V = C^{AB}(D_A D_B \alpha - \frac{1}{4} D_A \Delta V_B + \frac{1}{4}D_A V_B)
\ee
which exactly coincides with the charge given in \cite{cl2}.\footnote{Up to a total $u$ derivative term  $-\frac{1}{4}\alpha \dot{C}^{AB} C_{AB}$ that integrates to zero with the fall-offs underlaying the definition of these charges \cite{cl2}. }

The magnetic charge has the form (\ref{rhoV}) with $\E$ replaced by $\B$. Using  (\ref{B0uA}) it is given by: 
\be \label{rhotV}
\rho^*_V = - V^{B} \epsilon_B^{\phantom{A}A} \partial_u  \overset{(0)}{\E^u_{\ A}} + \alpha \partial_u (u \overset{(0)}{\B^u_{\ u}})  
\ee
With the fall-offs $C_{AB} = O(u^{-1-\epsilon})$ under consideration, the last term in (\ref{rhotV}) integrates to zero.  Let
\be
W^A := - V^{B} \epsilon_B^{\phantom{A}A},
\ee
then the `hard' and `soft' pieces of $\rho^*_V$ can be written as:
\be
\rho^{* \text{hard}}_V = W^A H_A +D_A W^A  M^{\text{hard}} , \quad \rho^{* \text{soft}}_V = W^A S_A +D_A W^A M^{\text{soft}}.
\ee
where we used Eq. (\ref{angmomasp}) and discarded total sphere divergences. Comparing with (\ref{rhoVhs}) we see that the expressions coincide up to total $u$ derivatives. Here however we face an obstacle: Whereas the boundary term vanishes in $\rho^{* \text{soft}}_V$, for $\rho^{* \text{hard}}$ it contains a divergent term: $\lim_{u \to - \infty} u M^{\text{hard}}(u)$. Thus, as it stands the `magnetic' charges are ill-defined (except for curl-free $V^A$). We hope to clarify this and other aspects of `magnetic' charges in the future.

\end{document}